\pdfoutput=1
 
\documentclass[PAPER,atlasdraft=true,cernpreprint,texlive=2018,UKenglish,LANGEDIT=true,LANGSHOW=false,texmf]{atlasdoc}
 
\usepackage[backend=biber]{atlaspackage}
\usepackage{atlasbiblatex}
 
\usepackage[process=true]{atlasphysics}
 
\newcommand\T{\rule{0pt}{2.6ex}}       
\newcommand\B{\rule[-1.2ex]{0pt}{0pt}} 
\newcommand{\hsp}{\hspace*{0.3cm}}
 
\newcolumntype{L}[1]{>{\raggedright\arraybackslash}p{#1}}
\newcolumntype{C}[1]{>{\centering\arraybackslash}p{#1}}
\newcolumntype{R}[1]{>{\raggedleft\arraybackslash}p{#1}}
 
\usepackage{afterpage}
\usepackage{multirow}
\usepackage{rotating}
\usepackage{comment}
\usepackage{pdflscape}
 
\usepackage{booktabs}
\usepackage{siunitx}
\usepackage{tablefootnote}
\graphicspath{{logos/}{figures/final/}}
 
\usepackage{ANA-HIGG-2018-51-PAPER-defs}

\AtlasTitle{Measurements of $WH$ and $ZH$ production in the $H \rightarrow b\bar{b}$ decay channel in $pp$ collisions at $13\,\text{Te\kern -0.1em V}$ with the ATLAS detector}

\AtlasAbstract{
Measurements of the Standard Model Higgs boson decaying into a $b\bar{b}$ pair and produced in association with a $W$ or $Z$ boson decaying into leptons, using proton--proton collision data collected between 2015 and 2018 by the ATLAS detector, are presented. The measurements use collisions produced by the Large Hadron Collider at a centre-of-mass energy of $\sqrt{s} = $13$\,\text{Te\kern -0.1em V}$, corresponding to an integrated luminosity of $139\,\mathrm{fb}^{-1}$.
The production of a Higgs boson in association with a $W$ or $Z$ boson is established with observed (expected) significances of 4.0 (4.1) and 5.3 (5.1) standard deviations, respectively.
Cross-sections of associated production of a Higgs boson decaying into bottom quark
pairs with an electroweak gauge boson, $W$ or $Z$, decaying into leptons are measured
as a function of the gauge boson transverse momentum in kinematic fiducial volumes.
The cross-section measurements are all consistent with the Standard Model expectations, and the total uncertainties vary
from 30\% in the high gauge boson transverse momentum regions to 85\% in the low regions. Limits are subsequently set on the parameters of an effective Lagrangian sensitive to modifications of the $WH$ and $ZH$ processes as well as the Higgs boson decay into $b\bar{b}$.
}
 
\author{The ATLAS Collaboration}
 
\AtlasRefCode{HIGG-2018-51}
 
\PreprintIdNumber{CERN-EP-2020-087}
 
\AtlasDate{\today}

\arXivId{1903.04618}
 
\HepDataRecord{hepdata.94800.v2}
 
\AtlasJournal{EPJC}
\AtlasJournalRef{Eur. Phys. J. C 81 (2021) 178}
\AtlasDOI{10.1140/epjc/s10052-020-08677-2}
\hypersetup{pdftitle={ATLAS document},pdfauthor={The ATLAS Collaboration}}

\begin{document}
 
\maketitle
 
\tableofcontents
 
\clearpage

\section{Introduction}
The Higgs boson~\autocite{Englert:1964et,Higgs:1964ia,Higgs:1964pj,Guralnik:1964eu,Higgs:1966ev,Kibble:1967sv} was
discovered in 2012 by the ATLAS and CMS
Collaborations~\autocite{HIGG-2012-27,CMS-HIG-12-028} with a mass of approximately 125\,\GeV\, from the analysis of
proton--proton ($pp$) collisions produced by the Large Hadron Collider (LHC)~\autocite{Evans:2008zzb}. Since then,
the analysis of data collected at centre-of-mass energies of 7\,\TeV, 8\,\TeV\ and 13\,\TeV\
in Runs 1 and 2 of the LHC has led to the observation and measurement of many of the production
modes and decay channels predicted by the Standard Model (SM)~\cite{HIGG-2014-14, HIGG-2016-21, HIGG-2016-22, HIGG-2016-25, HIGG-2016-33, HIGG-2017-11, HIGG-2016-07, CMS-HIG-17-025, CMS-HIG-16-040, CMS-HIG-16-041, CMS-HIG-16-042,HIGG-2015-07,HIGG-2017-07,CMS-HIG-16-043,HIGG-2018-13, CMS-HIG-17-035}.
 
The most likely decay mode of the SM Higgs boson is into pairs of $b$-quarks, with an expected
branching fraction of 58.2\% for a mass of $\mh=125$~\GeV~\autocite{Djouadi:1997yw, deFlorian:2016spz}. However, large
backgrounds from multi-jet production make a search in the dominant gluon--gluon fusion production mode very challenging at hadron colliders~\cite{Sirunyan:2017dgc}.
The most sensitive production modes for detecting \Hbb\ decays are the associated production of a Higgs
boson and a $W$ or $Z$ boson~\cite{Glashow:1978ab}, referred to as the $VH$ channel ($V$ = $W$ or $Z$), where the leptonic decay of the
vector boson enables efficient triggering and a significant reduction of the multi-jet
background.
As well as probing the dominant decay of the Higgs boson, this measurement allows the overall Higgs boson decay width~\cite{Lafaye:2009vr,Heinemeyer:2013tqa} to be constrained, provides the best sensitivity to the $WH$ and $ZH$ production modes and allows Higgs boson production at high transverse momentum to be probed, which provides enhanced sensitivity to some beyond the SM (BSM) physics models in effective field theories~\cite{Ellis:2014dva}.
The $\bbbar$ decay of the Higgs boson was observed by the ATLAS~\cite{HIGG-2018-04} and
CMS Collaborations~\cite{CMS-HIG-18-016} using data collected at centre-of-mass energies of 7\,\TeV, 8\,\TeV\ and 13\,\TeV\ during Runs 1 and 2 of the LHC. ATLAS also used the same dataset to perform differential measurements of the $VH$, $H \rightarrow b\bar{b}$ cross-section in kinematic fiducial volumes defined in the simplified template cross-section (STXS) framework~\cite{HIGG-2018-50}.
These measurements were used to set limits on the parameters of an effective Lagrangian sensitive to anomalous Higgs boson couplings with the electroweak gauge bosons.
 
This paper updates the measurements of the SM Higgs boson decaying into a $b\bar{b}$ pair in the
$VH$ production mode with the ATLAS detector in Run 2 of the LHC
presented in Refs.~\cite{HIGG-2018-04, HIGG-2018-50} and uses the full dataset. Events are categorised
in 0-, 1- and 2-lepton channels, based on the number of charged leptons, $\ell$ (electrons or
muons\footnote{This includes electrons and muons produced from the leptonic decay of a $\tau$-lepton.}), to explore the $ZH \to \nu\nu b\bar{b}$,  $WH \to \ell \nu b \bar{b}$ and $ZH \to\ell\ell b\bar{b}$ signatures, respectively. The dominant background processes after the
event selection are \vjets, \ttbar, single-top-quark and diboson production. Multivariate discriminants, built from
variables that describe the kinematics, jet flavour and missing transverse momentum content of the selected events, are used to maximise the sensitivity
to the Higgs boson signal. Their output distributions are used as inputs to a binned maximum-likelihood fit, referred to as the global likelihood fit,
which allows the yields and kinematics of both the signal and the background processes to be estimated. This method is validated using a diboson analysis, where the nominal multivariate analysis is modified to extract the $VZ$, $Z \rightarrow b\bar{b}$ diboson process. The Higgs boson signal measurement is also cross-checked with a dijet-mass analysis, where the signal yield is measured using the mass of the dijet system as the main observable instead of the multivariate discriminant. Finally, limits are set on the coefficients of effective Lagrangian operators which affect the $VH$ production and the $H \to b\bar{b}$ decay. Limits are reported for both the variation of a single operator and also the simultaneous variation of an orthogonal set of linear combinations of operators to which the analysis is sensitive.
 
This update uses \lumiFull\ of $pp$ collision data collected at a centre-of-mass energy of 13\,\TeV, to be compared with 79.8\,\ifb\ for the previous result. In addition, several improvements have been implemented: enhanced object calibrations, more coherent categorisation between the event selection and the STXS binning, re-optimised multivariate discriminants including the addition of more information, redefined signal and control regions, a significant increase in the effective number of simulated events and re-derived background modelling uncertainties, including using a multivariate approach to estimate the modelling uncertainty in the dominant backgrounds. A complementary analysis using the same final states, but focussing on regions of higher Higgs boson transverse momentum not accessible using the techniques outlined in this paper, has also been undertaken~\cite{ATLAS-CONF-2020-007}. The same dataset was used, resulting in some overlap in the events analysed.

\section{The ATLAS detector}
ATLAS~\cite{PERF-2007-01} is a general-purpose particle detector covering nearly the entire solid angle\footnote{ATLAS uses a right-handed coordinate system with its origin at the nominal interaction point (IP) in the centre of the detector and the $z$-axis coinciding with the axis of the beam pipe.  The $x$-axis points from the IP towards the centre of the LHC ring, and the $y$-axis points upward. Cylindrical coordinates ($r$,$\phi$) are used in the transverse plane, $\phi$ being the azimuthal angle around the $z$-axis. The pseudorapidity is defined in terms of the polar angle $\theta$ as $\eta = - \ln \tan(\theta/2)$. The distance in ($\eta$,$\phi$) coordinates, $\Delta R = \sqrt{(\Delta\phi)^2+(\Delta\eta)^2}$, is also used to define cone sizes. Rapidity is defined as $y=(1/2) \ln[(E+p_z)/(E-p_z)]$, where $E$ is the energy and $p_z$ is the $z$-component of the momentum. Transverse momentum and energy are defined as $\pt=p\sin\theta$ and $\et=E\sin\theta$, respectively.}
around the collision point.  An inner tracking detector, located within a 2~T axial
magnetic field generated by a thin superconducting solenoid, is used to measure the
trajectories and momenta of charged particles. The inner layers consist of
high-granularity silicon pixel detectors covering a pseudorapidity
range $|\eta| < 2.5$, with an innermost layer~\cite{Capeans:1291633,Abbott:2018ikt} that was added to the detector between Run 1 and Run 2.
Silicon microstrip detectors covering $|\eta| < 2.5$ are located beyond the pixel detectors.
Outside the microstrip detectors and covering $|\eta | < 2.0$, there are straw-tube tracking detectors, which also provide measurements of transition radiation that are used in electron identification.
 
A calorimeter system surrounds the inner tracking detector, covering the pseudorapidity range $|\eta| < 4.9$.
Within the region $|\eta|< 3.2$, electromagnetic calorimetry is provided by barrel ($|\eta| < 1.475$) and
endcap ($1.375 < |\eta| < 3.2$) high-granularity lead/liquid-argon (LAr) sampling calorimeters,
with an additional thin LAr presampler covering $|\eta| < 1.8$ to correct for energy loss in material upstream of the calorimeters.
Hadronic calorimetry is provided by a steel/scintillator-tile calorimeter within $|\eta| < 1.7$, and copper/LAr endcap calorimeters extend the coverage to $|\eta|=3.2$.
The solid angle coverage for $|\eta|$ between 3.2 and 4.9 is completed with copper/LAr and tungsten/LAr calorimeter modules optimised for electromagnetic and hadronic measurements, respectively.
 
The outermost part of the detector is the muon spectrometer, which measures the curved trajectories of muons in the magnetic field of three large air-core superconducting toroidal magnets. High-precision tracking is performed within the range $|\eta| < 2.7$ and there are chambers for fast triggering within the range $|\eta| < 2.4$.
 
A two-level trigger system~\cite{Aaboud:2016leb} is used to reduce the recorded data rate.
The first level is a hardware implementation aiming to reduce the rate to around
100~kHz, while the software-based high-level trigger provides the remaining rate reduction to approximately 1~kHz.

 
\section{Data and simulated event samples}
\label{sec:samples}
The data used in this analysis were collected using unprescaled single-lepton or missing transverse momentum triggers at a centre-of-mass energy of 13~\TeV\
during the 2015--2018 running periods.
Events are selected for analysis only if they are of good quality and if all the relevant
detector components are known to have been in good operating condition, which corresponds to a total integrated luminosity of
$139.0\pm 2.4$~\fb~\cite{DAPR-2013-01,Avoni:2018iuv}.
The recorded events contain an average of 34 inelastic $pp$ collisions per bunch-crossing.
 
Monte Carlo (MC) simulated events are used to model most of the backgrounds from SM processes and the $VH$, $H
\rightarrow b\bar{b}$ signal processes. A summary of all the generators used for the simulation of the signal and background
processes is shown in Table~\ref{tab:samples}. Samples produced with alternative generators are used to estimate systematic uncertainties in the event modelling, as described in Section~\ref{sec:syst}. The same event generators as in Ref.~\cite{HIGG-2018-04} are used; however, the number of simulated events in all samples has been increased by at least the factor by which the integrated luminosity grew compared to the previous publication ($\sim1.75$). In addition, processes which significantly contributed to the statistical uncertainty of the background in the previous publication benefited from a further factor of two increase in the number of simulated events produced.
 
All simulated processes are normalised using the most accurate theoretical cross-section predictions currently available and were generated
to next-to-leading-order (NLO) accuracy at least, except for the $gg  \to ZH$ and $gg  \to VV$ processes, which were generated at LO. All samples of simulated events were passed through the ATLAS detector simulation~\cite{SOFT-2010-01} based on $\GEANT$~\cite{geant}. The effects
of multiple interactions in the same and nearby bunch crossings (pile-up) were modelled by overlaying minimum-bias events, simulated using the soft QCD processes of \textsc{Pythia 8.186}~\cite{Pythia8} with the A3~\cite{ATL-PHYS-PUB-2016-017} set of tuned parameters (tune) and \textsc{NNPDF2.3LO}~\cite{Ball:2012cx} parton distribution functions (PDF). For all samples of simulated events, except for those generated using \textsc{Sherpa}~\cite{sherpa}, the \textsc{EvtGen v1.6.0} program~\cite{Lange:2001uf} was used to describe the decays of bottom and charm hadrons.
 
\begin{table}[hbt!]
\begin{center}{\fontsize{7}{7.9}\selectfont
\caption
{The generators used for the simulation of the signal and background
processes. Samples are generated considering decays into all three lepton ($\ell$) flavours. If not specified, the order of the cross-section calculation refers to the expansion in the strong coupling constant ($\alphas$).
The acronyms ME, PS and UE stand for matrix element, parton shower and underlying event,  respectively.
$(\star)$ The events were generated using the first PDF in the NNPDF3.0NLO set and
subsequently reweighted to the PDF4LHC15NLO set~\cite{Butterworth:2015oua} using the internal algorithm in \textsc{Powheg-Box v2}.
$(\dagger)$ The NNLO(QCD)+NLO(EW) cross-section calculation for the $pp \to ZH$ process already includes the $gg\to ZH$ contribution.
The $qq\to ZH$  process is normalised using the cross-section for the $pp \to ZH$ process,
after subtracting the $gg\to ZH$ contribution. An additional scale factor is applied to
the $qq \rightarrow VH$ processes as a function of the transverse momentum of the vector
boson, to account for electroweak (EW) corrections at NLO\@. This makes use of the $VH$ differential cross-section computed with \textsc{Hawk}~\cite{Denner:2011id,Denner:2014cla}. Contributions from photon-induced processes are also included for $pp \to WH$~\cite{deFavereaudeJeneret:2009db}. ($\ddagger$) For the diboson samples the cross-sections are calculated by the Monte Carlo generator at NLO accuracy in QCD.
\protect\label{tab:samples}}
\begin{tabular}{llllll}
\toprule
\hsp Process & ME generator & ME PDF &PS and & UE model & Cross-section \hspace{2.5cm}\\
& & & Hadronisation & tune & order\\
\midrule
\multicolumn{6}{l}{Signal, mass set to 125~\GeV\ and $b\bar{b}$ branching fraction to 58\%} \\
\midrule
\hsp $qq\to WH$  &\textsc{Powheg-Box v2}~\cite{Alioli:2010xd} +&
NNPDF3.0NLO$^{(\star)}$~\cite{Ball:2014uwa} &\PYTHIAV{8.212}~\cite{Pythia8} & AZNLO~\cite{AZNLO:2014} & NNLO(QCD)$^{(\dagger)}$+  \\
\hspace*{0.5cm} $\to\ell\nu b\bar{b}$&   \textsc{GoSam}~\cite{arXiv:1111.2034} + \textsc{MiNLO}~\cite{arXiv:1206.3572,Luisoni:2013kna}  & & & &NLO(EW)~\cite{Ciccolini:2003jy,Brein:2003wg,Ferrera:2011bk,Brein:2011vx,Ferrera:2013yga,Ferrera:2014lca,Campbell:2016jau} \\
\hsp $qq\to ZH$ &\textsc{Powheg-Box v2} + & NNPDF3.0NLO$^{(\star)}$ &\PYTHIAV{8.212} & AZNLO & NNLO(QCD)$^{(\dagger)}$+  \\
\hspace*{0.5cm}$\to\nu\nu b\bar{b}/\ell\ell b\bar{b}$&   \textsc{GoSam} + \textsc{MiNLO}  & & & &NLO(EW) \\
\hsp $gg  \to ZH$ &\textsc{Powheg-Box v2} & NNPDF3.0NLO$^{(\star)}$ &\PYTHIAV{8.212} & AZNLO & NLO+  \\
\hspace*{0.5cm}$\to\nu\nu b\bar{b}/\ell\ell b\bar{b}$&  & & & &NLL~\cite{Altenkamp:2012sx,Hespel:2015zea,ggzhnll,Harlander:2013mla,Brein:2012ne}\\
\midrule
\multicolumn{6}{l}{Top quark, mass set to 172.5~\GeV}  \\
\midrule
\hsp $t\bar{t}$ &\textsc{Powheg-Box v2}~\cite{Frixione:2007nw} &  NNPDF3.0NLO &\PYTHIAV{8.230} &A14~\cite{ATL-PHYS-PUB-2014-021}& NNLO+NNLL~\cite{Czakon:2011xx} \\
\hsp $s$-channel single top&\textsc{Powheg-Box v2}~\cite{Alioli:2009je} & NNPDF3.0NLO &\PYTHIAV{8.230} &A14 & NLO~\cite{PhysRevD.81.054028} \\
\hsp $t$-channel single top&\textsc{Powheg-Box v2}~\cite{Alioli:2009je} & NNPDF3.0NLO &\PYTHIAV{8.230} &A14& NLO~\cite{Kidonakis:2011wy} \\
\hsp $Wt$ &\textsc{Powheg-Box v2}~\cite{Re:2010bp} & NNPDF3.0NLO &\PYTHIAV{8.230} &A14& Approximate NNLO~\cite{PhysRevD.82.054018} \\
\midrule
\multicolumn{6}{l}{Vector boson + jets} \\
\midrule
\hsp $W\to\ell\nu$ & \textsc{Sherpa 2.2.1}~\cite{sherpa,Cascioli:2011va,Gleisberg:2008fv} & NNPDF3.0NNLO &  \textsc{Sherpa 2.2.1}~\cite{Schumann:2007mg,Hoeche:2012yf} & Default & NNLO~\cite{Catani:2009sm}  \\
\hsp $Z/\gamma^{*}\to\ell\ell$ & \textsc{Sherpa 2.2.1} & NNPDF3.0NNLO &  \textsc{Sherpa 2.2.1} & Default & NNLO    \\
\hsp $Z\to\nu\nu$ & \textsc{Sherpa 2.2.1} & NNPDF3.0NNLO &  \textsc{Sherpa 2.2.1} & Default & NNLO    \\
\midrule
\multicolumn{6}{l}{Diboson} \\
\midrule
\hsp $qq \to WW$ & \textsc{Sherpa 2.2.1} & NNPDF3.0NNLO &  \textsc{Sherpa 2.2.1} & Default & NLO$^{(\ddagger)}$  \\
\hsp $qq \to WZ$ & \textsc{Sherpa 2.2.1} & NNPDF3.0NNLO &  \textsc{Sherpa 2.2.1} & Default & NLO$^{(\ddagger)}$  \\
\hsp $qq \to ZZ$ & \textsc{Sherpa 2.2.1} & NNPDF3.0NNLO &  \textsc{Sherpa 2.2.1} & Default & NLO$^{(\ddagger)}$  \\
\hsp $gg \to VV$ & \textsc{Sherpa 2.2.2} & NNPDF3.0NNLO &  \textsc{Sherpa 2.2.2} & Default & NLO$^{(\ddagger)}$  \\
\bottomrule
\end{tabular}}
\end{center}
\end{table}
 
\section{Object and event selection}
The event topologies characteristic of $VH$,~$H\to\bbbar$ processes contain zero, one or
two charged leptons, and two `$b$-jets' containing particles from $b$-hadron decays. The
object and event selections broadly follow those of
Ref.~\cite{HIGG-2018-04} but with updates to the definition of the signal and control regions.
 
\subsection{Object reconstruction}
 
Tracks measured in the inner detector are used to reconstruct interaction
vertices~\cite{Aaboud:2016rmg}, of which the one with the highest sum of squared
transverse momenta of associated tracks is selected as the primary vertex of the hard interaction.
 
Electrons are reconstructed from topological clusters of energy deposits in the electromagnetic calorimeter and matched to a track in the
inner detector~\cite{EGAM-2018-01}. Following Refs.~\cite{EGAM-2018-01,HIGG-2016-29}, \textit{loose} electrons are required
to have $\pt>7\,\GeV$ and $\abseta<2.47$, to have small impact
parameters,\footnote{Transverse and longitudinal impact parameters are defined relative
to the primary vertex position, where the beam line is used to approximate the primary
vertex position in the transverse plane.} to fulfil a loose track isolation requirement,
and to meet a `LooseLH' quality criterion computed from shower shape,
track quality and track--cluster matching variables. In the 1-lepton channel, \textit{tight}
electrons are selected using a `TightLH' likelihood requirement and a
calorimeter-based isolation in addition to the track-based isolation.
 
Muons are required to be within the acceptance of the muon spectrometer $\abseta<2.7$, to
have $\pt>7\,\GeV$, and to have small impact parameters. \textit{Loose} muons are selected
using a `loose' quality criterion~\cite{PERF-2015-10} and a loose track
isolation requirement. In the 1-lepton channel, \textit{tight} muons fulfil the `medium' quality
criterion and a stricter track isolation requirement.
 
Hadronically decaying $\tau$-leptons~\autocite{PERF-2014-06,ATLAS-CONF-2017-029} are
required to have $\pt>20\,\GeV$ and $\abseta<2.5$, to be outside the transition region
between the barrel and endcap electromagnetic calorimeters $1.37<\abseta<1.52$, and to meet a `medium' quality
criterion~\cite{ATLAS-CONF-2017-029}. Reconstructed hadronic $\tau$-leptons are not directly used in the event selection, but are utilised in the missing transverse momentum calculation and are also used to avoid double-counting hadronic $\tau$-leptons as other objects.
 
Jets are reconstructed from the energy in topological clusters of calorimeter cells~\cite{ATL-LARG-PUB-2008-002}
using the \antikt\ algorithm~\cite{Cacciari:2008gp} with radius
parameter $R=0.4$.  Jet cleaning criteria are used to identify jets arising from non-collision backgrounds or noise in the
calorimeters~\cite{ATLAS-CONF-2015-029}, and events containing such jets are removed.
Jets are required to have $\pt>20\,\GeV$ in the central region
($\abseta<2.5$), and $\pt>30\,\GeV$ outside the tracker acceptance ($2.5<\abseta<4.5$).
A jet vertex tagger~\cite{Aad:2015ina} is used to remove jets with $\pt<120\,\GeV$ and $\abseta<2.5$ that are identified as not being associated with the primary vertex of the hard interaction.
Simulated jets are labelled as $b$-, $c$- or light-flavour jets according to which hadrons with
$\pt>5\,\GeV$ are found within a cone of size $\Delta R=0.3$ around their axis~\cite{FTAG-2018-01}.
In the central region, jets are identified as $b$-jets ($b$-tagged) using a multivariate discriminant~\autocite{FTAG-2018-01} (MV2), with the selection tuned to produce an average efficiency
of 70\% for $b$-jets in simulated $\ttbar$ events, which corresponds to light-flavour ($u$-, $d$-, $s$-quark and gluon) jet and $c$-jet misidentification efficiencies of 0.3\% and 12.5\% respectively.
 
Simulated $V$+jets events are categorised according to the two $b$-tagged jets
that are required in the event: $V+ll$ when they are both light-flavour jets, $V+cl$ when there is one
$c$-jet and one light-flavour jet, and \vhf\ (heavy flavour) in all other cases (which after the $b$-tagging selection mainly consist of events with two $b$-jets).
 
In practice, $b$-tagging is not applied directly to simulated events containing light-flavour jets or $c$-jets, because the substantial MV2 rejection results in a significant statistical uncertainty for these background processes. Instead, all events with $c$-jets or light-flavour jets are weighted by the probability that these jets pass the $b$-tagging requirement~\cite{HIGG-2016-29}. This is an expansion of the weighting technique compared to the previous analysis, where only jets in the $V+ll$, $V+cl$ and $WW$ processes were treated in this manner. Applying the same treatment to all light-flavour jets and $c$-jets significantly increases the number of simulated events present after the full event selection, reducing the statistical uncertainty of the \vhf\ ($\ttbar$) background by $\sim65\%$--$75\%$ ($\sim25\%$).  When comparing the direct application of the $b$-tagging to the weighting technique, differences were observed in a particular subset of events with a small angular separation between the jets, but it was verified that this has a negligible impact on the result.
 
In addition to the standard jet energy scale calibration~\autocite{PERF-2016-04}, $b$-tagged jets receive additional flavour-specific corrections to improve their energy measurement (scale and resolution): if any muons are found within a \pT-dependent cone around the jet axis, the four-momentum of the closest muon is added to that of the jet. In addition, a residual correction is applied to equalise the
response to jets with leptonic or hadronic decays of heavy-flavour hadrons and to correct for resolution effects. This improves the resolution of the dijet mass by up to $\sim20\%$~\cite{HIGG-2016-29}. Alternatively, in the
2-lepton channel for events with two or three jets, a per-event kinematic likelihood uses the complete reconstruction of all final-state objects
to improve the estimate of the energy of the $b$-jets. This improves the resolution of the dijet mass by up to $\sim40\%$.
 
The missing transverse momentum, $\vecmet$, is reconstructed as the negative vector sum of the transverse
momenta of leptons, photons, hadronically decaying $\tau$-leptons and jets, and a `soft-term', $\vec{p}_{\mathrm{T}}^{\mathrm{miss},\mathrm{st}}$. The soft-term is calculated as the vectorial sum of the \pt\ of tracks matched to the primary vertex but not associated with a reconstructed lepton or jet~\autocite{PERF-2016-07}. The
magnitude of $\vecmet$ is referred to as \met. The track-based missing transverse momentum, $\vec{p}_{\mathrm{T}}^{\mathrm{miss}}$, is calculated using only tracks reconstructed in the inner tracking detector and matched to the primary vertex.
 
An overlap removal procedure is applied to avoid any double-counting between leptons, including hadronically decaying $\tau$-leptons, and jets.
 
\subsection{Event selection and categorisation}
 
Events are categorised into 0-, 1- and 2-lepton channels (referred to as the $n$-lepton channels) depending on the number of
selected electrons and muons, to target the $ZH \to vv b\bar{b}$, $WH \to \ell \nu b \bar{b}$
and $ZH \to \ell\ell b\bar{b}$ signatures, respectively. In all channels, events are
required to have exactly two $b$-tagged jets, which form the Higgs boson candidate. At least one
$b$-tagged jet is required to have \pt\ greater than 45\,\GeV.
Events are further split into 2-jet or 3-jet categories, where the 3-jet category includes events with one or more untagged jets. In the 0- and 1-lepton channels, only one untagged jet is allowed, as the \ttbar\ background is much larger in events with four jets or more. In the 2-lepton channel any number of untagged jets are accepted in the 3-jet category (referred to as the $\ge$ 3-jet category when discussing only the 2-lepton channel), which increases the signal acceptance in this category by 100\%.

The reconstructed transverse momentum of the vector boson, \ptv, corresponds to \met\ in the 0-lepton
channel, the vectorial sum of $\vecmet$ and the charged-lepton transverse momentum in the
1-lepton channel, and the transverse momentum of the 2-lepton system in the 2-lepton
channel. Since the signal-to-background ratio increases for large \ptv\ values~\autocite{Butterworth:2008iy,ATL-PHYS-PUB-2009-088},
the analysis focuses on two high-\ptv\ regions defined as $150\,\GeV<\ptv<250\,\GeV$ and $\ptv>250\,\GeV$. In the 2-lepton
channel, an additional fiducial measurement region is studied via the inclusion of a medium-\ptv\ region
with $75\,\GeV<\ptv<150\,\GeV$.
 
The event selection for the three lepton channels is outlined in Table~\ref{tbl:evselTable} with details provided below.
 
\afterpage{
\begin{landscape}
\begin{table}[ptbh]
\centering
\caption{Summary of the event selection and categorisation in the 0-, 1- and 2-lepton channels.}
\label{tbl:evselTable}
\begin{tabular}{l c c c c}
\toprule
\multirow{2}{*}{Selection} & 0-lepton & \multicolumn{2}{c}{1-lepton}        & 2-lepton\\
&          & $e$ sub-channel & $\mu$ sub-channel &         \\
\midrule
Trigger                      & \met\    & Single lepton   & \met\             & Single lepton \B \\
\midrule
\multirow{4}{*}{Leptons}     & \multirow{4}{*}{0 \textit{loose} leptons}   & \multirow{2}{*}{Exactly 1 \textit{tight} electron} & \multirow{2}{*}{Exactly 1 \textit{tight} muon}    & Exactly 2 \textit{loose}  leptons \\
&                            & \multirow{2}{*}{0 additional \textit{loose} leptons} &  \multirow{2}{*}{0 additional \textit{loose} leptons}  & $\pt>27$~\GeV \\
&                            & \multirow{2}{*}{$\pt>27$~\GeV}                       & \multirow{2}{*}{$\pt>25$~\GeV}                         & Same-flavour  \\
&                            &                                     &                                       & Opposite-sign charges ($\mu\mu$) \\
\met\                        & $>$ 150~\textrm{\GeV}      & $>$ 30~\textrm{\GeV}                & --                                    & --  \\
\mll\                        & --                         & --                                  & --                                    & 81~\textrm{\GeV}~$<$ \mll\ $<$ 101~\textrm{\GeV} \\
\midrule
\multirow{2}{*}{Jet \pt}     & \multicolumn{4}{c}{$>$ 20 \textrm{\GeV} for $|\eta|$~$<$~2.5} \\
& \multicolumn{4}{c}{$>$ 30 \textrm{\GeV} for 2.5~$<$~$|\eta|$~$<$~4.5} \\
$b$-jets                     & \multicolumn{4}{c}{Exactly 2 $b$-tagged jets} \\
Leading $b$-tagged jet \pt\  & \multicolumn{4}{c}{$>$ 45 \textrm{\GeV}} \\
\midrule
Jet  categories                         & Exactly 2 / Exactly 3 jets & \multicolumn{2}{c}{Exactly 2 / Exactly 3 jets}                                                           & Exactly 2 / $\geq$ 3 jets \\
\midrule
$H_{\mathrm{T}}$                                                & $>$ 120 \textrm{\GeV} (2 jets), $>$150 \textrm{\GeV} (3 jets) & \multicolumn{2}{c}{--}  & -- \\
$\textrm{min}[\Delta\phi(\vecmet,\vec{\mathrm{jets}}) ]$      & $> 20\ensuremath{^\circ}$ (2 jets), $> 30\ensuremath{^\circ}$ (3 jets) & \multicolumn{2}{c}{--}  & -- \\
$\Delta\phi(\vecmet, \vec{bb})$                               & $> 120\ensuremath{^\circ}$ & \multicolumn{2}{c}{--}   & -- \\
$\Delta\phi(\vec{b_1},\vec{b_2})$ & $< 140\ensuremath{^\circ}$ & \multicolumn{2}{c}{--}                               & -- \\
$\Delta\phi(\vecmet,\vec{p}_{\mathrm{T}}^{\mathrm{miss}})$        & $< 90\ensuremath{^\circ}$  & \multicolumn{2}{c}{--}   & -- \\
\midrule
\multirow{3}{*}{$\pt^V$ regions}                              & --                        & \multicolumn{2}{c}{--}    & 75~\textrm{\GeV}~$<$ $\pt^V$ $<$ 150~\textrm{\GeV} \T \B \\
& 150~\textrm{\GeV}~$<$ $\pt^V$ $<$ 250~\textrm{\GeV} \T \B & \multicolumn{2}{c}{150~\textrm{\GeV}~$<$ $\pt^V$ $<$ 250~\textrm{\GeV}} \T \B & 150~\textrm{\GeV}~$<$ $\pt^V$ $<$ 250~\textrm{\GeV} \T \B  \\
& $\pt^V$ $>$ 250~\textrm{\GeV} \T \B  & \multicolumn{2}{c}{$\pt^V$ $>$ 250~\textrm{\GeV}} \T \B & $\pt^V$ $>$ 250~\textrm{\GeV} \T \B \\
\midrule
Signal regions              & \multicolumn{4}{c}{$\Delta R(\vec{b_1},\vec{b_2})$ signal selection}  \\
\midrule
Control regions             & \multicolumn{4}{c}{High and low $\Delta R(\vec{b_1},\vec{b_2})$ side-bands}  \\
\bottomrule
\end{tabular}
\end{table}
\end{landscape}
}
 
\paragraph{0-lepton channel} The online selection uses \met\ triggers with
thresholds that varied from 70\,\GeV\ to 110\,\GeV\ between the 2015
and 2018 data-taking periods. Their efficiency is measured in $W$+jets, $Z$+jets and $\ttbar$ events
using single-muon triggered data, which effectively
selects events with large trigger-level \met\ values as muons are not included in the trigger \met\ calculation. The resulting trigger correction factors that are applied to
the simulated events range from 0.95 at the offline \met\ threshold of 150\,\GeV\ to a
negligible deviation from unity at \met\ values above 200\,\GeV.
A requirement on the scalar sum of the transverse momenta of the jets, $H_\mathrm{T}$, removes a
small part of the phase space (less than 1\%) where the trigger efficiency depends mildly on the number of
jets in the event. Events with any \textit{loose} lepton are rejected. High \met\ in multi-jet
events typically arises from mismeasured jets in the calorimeters. Such events are
efficiently removed by requirements on the angular
separation of the $\vecmet$, jets, and $\vec{p}_{\mathrm{T}}^{\mathrm{miss}}$.
 
\paragraph{1-lepton channel} In the electron sub-channel, events are required to satisfy a logical
OR of single-electron triggers with \pt\ thresholds that started at 24\,\GeV\ in 2015
and increased to 26\,\GeV\ in 2016--2018.\footnote{Additional identification and isolation requirements are applied to the trigger object to allow a low \pt\ threshold to be maintained throughout Run 2.}  The muon sub-channel uses the same \met\
triggers and correction factors as the 0-lepton channel. As these triggers effectively
select on \ptv, given that muons are not included in the trigger \met\ calculation, they perform more efficiently than the single-muon triggers in the analysis phase
space, which have a lower efficiency due to the more limited coverage of the muon trigger system in the central region. Events are required to have exactly one \textit{tight} muon with \pt\ > 25\,\GeV\ or one \textit{tight} electron with \pt\ > 27\,\GeV\ and no additional \textit{loose} leptons. In the electron sub-channel an additional
selection of $\met>30\,\GeV$ is applied to reduce the background from multi-jet
production.

\paragraph{2-lepton channel} The trigger selection in the electron sub-channel is the same
as in the 1-lepton channel. In the muon sub-channel, an OR of single-muon triggers
is used, with lowest \pt\ thresholds increasing from 2016--2018 and ranging from 20\,\GeV\ to 26\,\GeV.
Events must have exactly two same-flavour \textit{loose} leptons, one of which must have
$\pt>27\,\GeV$, and the invariant mass of the lepton pair must be close to the $Z$ boson mass.
In dimuon events, the two muons are required to have opposite-sign charge. This is not used in the electron sub-channel, where the charge misidentification rate is not negligible.
 
\paragraph{Signal and control regions} The three $n$-lepton channels, two jet categories and two (0-lepton, 1-lepton) or three (2-lepton) \ptv\ regions result in a total of 14 analysis regions. Each analysis region is further split into a signal region (SR) and two control regions (CRs), resulting in a total of 42 regions. The CRs are enriched in either \vhf\ or \ttbar\ events and defined using a continuous selection on the $\Delta R$ between the two
$b$-tagged jets, $\Delta R(\vec{b}_1,\vec{b}_2)$, as a function of \ptv, with the $b$-tagged jets labelled in decreasing \pt\ as $b_1$ and $b_2$.
A lower and upper requirement on $\Delta R(\vec{b}_1,\vec{b}_2)$ is applied, creating two CRs, referred to as the low and high $\Delta R$ CRs, shown in Figure~\ref{fig:dR_vpt_selection}.
In the 1-lepton channel, the high $\Delta R$ selection was tuned such that the SR and low $\Delta R$ CR contain  95\% (85\%) of the signal in the 2-jet (3-jet) categories, whilst the low $\Delta R$ selection was tuned such that the SR contains 90\% of the diboson yield, to ensure that a sufficient number of these events remain when conducting the diboson validation analysis.
The same $\Delta R$ selection is applied in all three $n$-lepton channels and keeps over 93\% of the signal in the 2-jet categories and over 81\% (68\%) of the signal in the 3-jet ($\geq$ 3-jet) categories.\footnote{Although the higher jet multiplicity categories have a lower signal efficiency than the 2-jet categories, any reduction in the sensitivity in these categories is less than 5\%.}
\\
\begin{figure}[htbp]
\centering
\begin{tabular}{cccc}
\includegraphics[width=0.7\linewidth]{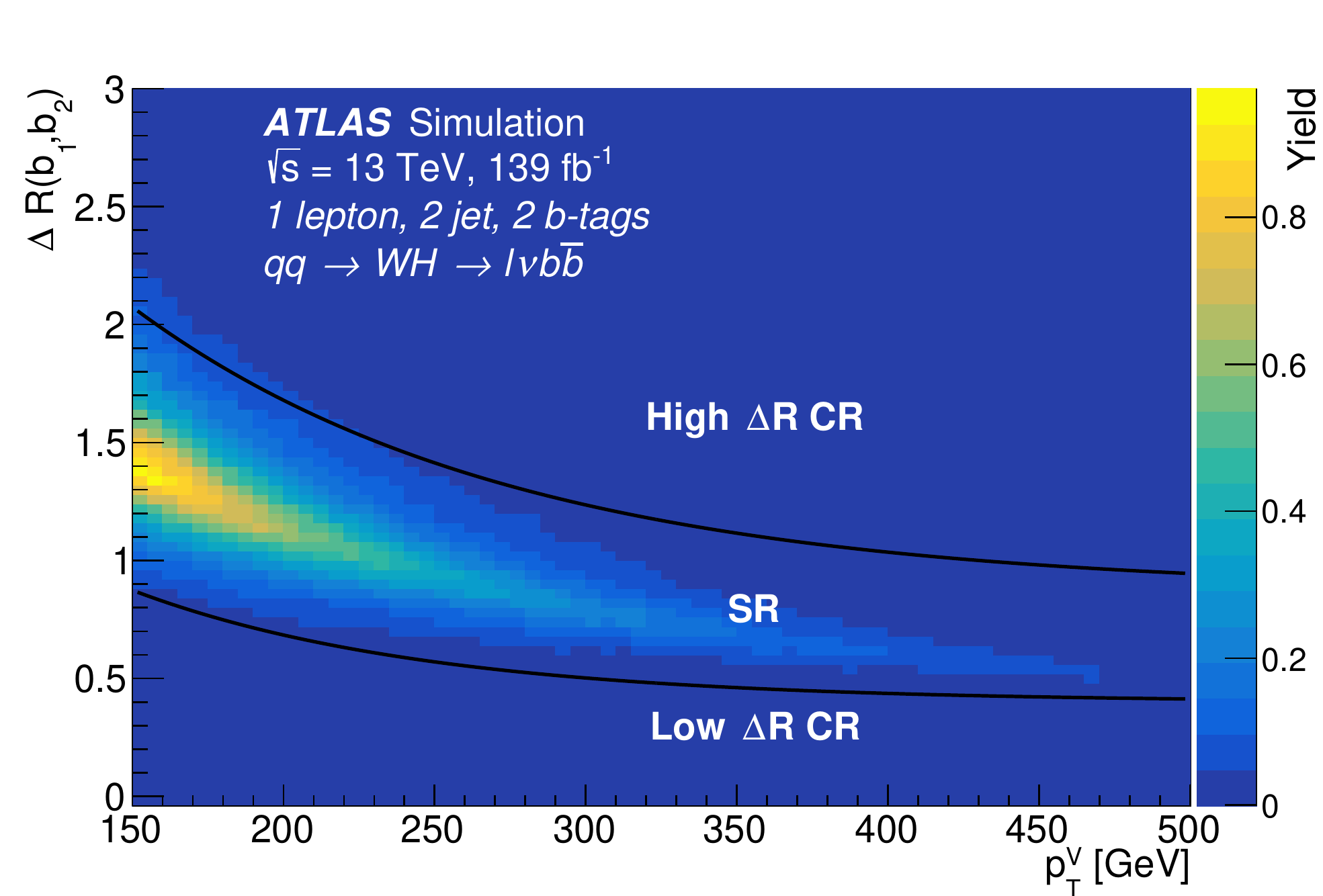} \\
\includegraphics[width=0.7\linewidth]{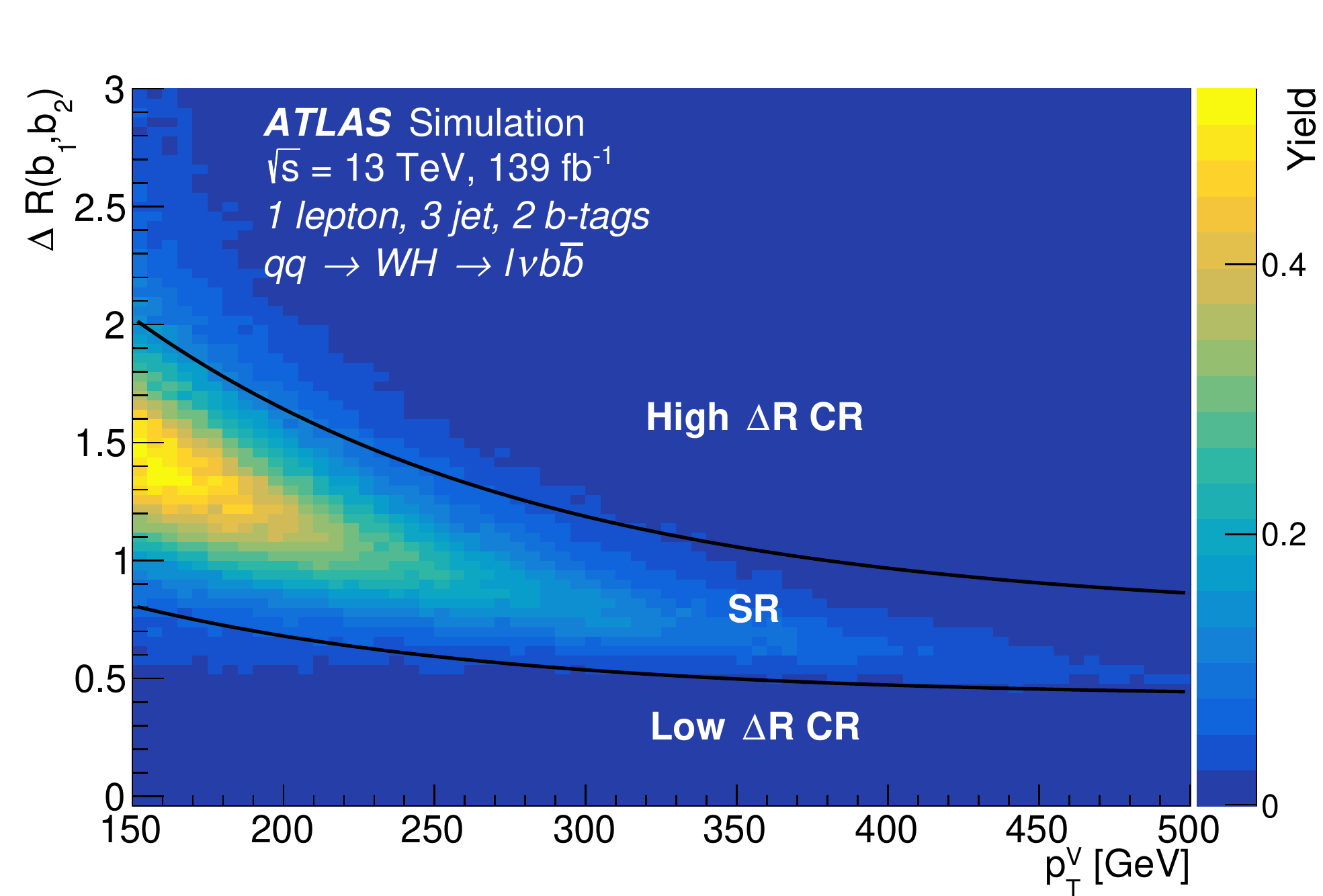}
\\
\end{tabular}
\caption{The signal yield distribution of the $\Delta R$ between the two
$b$-tagged jets, $\Delta R(\vec{b}_1,\vec{b}_2)$, as a function of \ptv\ in the 1-lepton channel for 2-$b$-tag events, in the 2-jet~(top) and exactly 3-jet (bottom) categories in the high-\ptv\ region. The lines demonstrate the continuous lower and upper selection on $\Delta R(\vec{b}_1,\vec{b}_2)$ used to categorise the events into the signal and control regions.
\label{fig:dR_vpt_selection}}
\end{figure}
 
The acceptances in the three $n$-lepton channels after the event selection, as well as the predicted
cross-sections times branching fractions for \wzh\ with \wln, \zll, \znn, and $H\to\bbbar$
are given in Table~\ref{tab:crosssection_acc}. The non-negligible acceptance for the $qq\to
WH$ process in the 0-lepton channel is mostly due to events with a hadronically decaying
$\tau$-lepton produced in the $W$ decay, which are not explicitly vetoed and which could also be misidentified as a jet or subsequently decay to a low-\pt\ electron or muon that fails to satisfy the selection criteria. The larger acceptance for the $gg\to ZH$
process compared with $qq\to ZH$  is due to the harder \ptv\ spectrum of the gluon-induced
process.
 
\begin{table}[tb!]
\begin{center}
\caption{The cross-section ($\sigma$) times branching fraction ($B$) and acceptance obtained from the simulated signal samples for the three channels
at $\sqrt{s}=13$~\TeV.
The $qq$- and $gg$-initiated $ZH$ processes are shown separately.
The branching fractions are calculated considering only decays into muons and electrons for \zll\
and decays into all three lepton flavours for \wln.
The acceptance is calculated as the fraction of events remaining in the combined signal and control
regions after the full event selection.
\label{tab:crosssection_acc}}
\begin{tabular}{lSSSSS}
\toprule
\multirow{2}{*}{Process} &
{\multirow{2}{*}{$\sigma$ $\times$ $B$ [fb]}} &
\multicolumn{3}{c}{Acceptance [\%]}\\
\cmidrule(lr){3-5}
&  & {0-lepton} & {1-lepton} & {2-lepton}\\
\midrule
$qq\to ZH\to\ell\ell b\bar{b}$  &  29.9  & <0.1     &  0.2 & 6.4 \\
$gg\to ZH\to\ell\ell b\bar{b}$   &   4.8  & <0.1      & 0.3 & 14.5 \\
$qq\to WH\to\ell\nu b\bar{b}$  & 269.0  & 0.2 & 1.1 & {--} \\
$qq\to ZH\to\nu\nu b\bar{b}$ &  89.1  & 1.9 & {--}        & {--} \\
$gg\to ZH\to\nu\nu b\bar{b}$   &   14.3  & 3.5 & {--}        & {--} \\
\bottomrule
\end{tabular}
\end{center}
\end{table}
 
\subsection{Simplified template cross-section categories}
 
Cross-section measurements are conducted in the reduced $VH$, $V\to$~leptons
stage-1.2 STXS region scheme~\cite{Badger:2016bpw, Berger:2019wnu} described in Ref.~\cite{HIGG-2018-50} and summarised in Table~\ref{tab:STXS_reducedstage1}. In this scheme, $qq\rightarrow ZH$ and $gg\rightarrow ZH$ are treated as a single $ZH$ process, since there is currently not enough sensitivity to distinguish between them. The expected signal distributions and acceptance times efficiencies for each STXS region are estimated from the simulated signal samples by selecting events using information from the generator's `truth' record, in particular the truth \ptv, denoted by $p_{\textrm{T}}^{V\textrm{, t}}$.
The signal yield in each reconstructed-event category
for each STXS region is shown in Figure~\ref{fig:5POI_acceptance}, with the corresponding
fraction of signal events shown in Figure~\ref{fig:5POI_purity}. The key improvement compared to the previous publication is the addition of a reconstructed-event category with $\pT^V>250$~\GeV. This region is more aligned with the STXS regions and significantly reduces the correlation between the STXS measurements in the two highest $\pT^{V\textrm{, t}}$ bins.
The acceptance times efficiency for $WH$ events with
$p_{\textrm{T}}^{W\textrm{, t}}<150$~\GeV\ or $ZH$ events
with $p_{\textrm{T}}^{Z\textrm{, t}}<75$~\GeV\ is at the level of 0.1\% or smaller.
Given the lack of sensitivity to these regions, the signal cross-section in these regions is
constrained to the SM prediction, within the theoretical
uncertainties. These regions contribute only marginally to the
selected event sample and the impact on the final results is negligible.
 
\begin{table}[!h]
\caption{
The simplified template cross-section regions used for measurements and the corresponding
reconstructed analysis regions that are most sensitive.
The current analysis is not sensitive to the regions $WH$, $p_{\textrm{T}}^{W\textrm{, t}}<150$~\GeV\ and $ZH$, $p_{\textrm{T}}^{Z\textrm{, t}}<75$~\GeV, and their
cross-sections are fixed to the SM prediction within their theoretical uncertainties. All leptonic decays of the weak gauge bosons (including $Z\to\tau\tau$ and $W\to\tau\nu$, which are extrapolated from the electron and muon channel measurements) are considered for the STXS definition.
}
\label{tab:STXS_reducedstage1}
\begin{center}
\setlength{\extrarowheight}{4pt}
\begin{tabular}{ccccc}
\toprule
\multicolumn{2}{c}{STXS region}                               &  \multicolumn{3}{c}{Corresponding reconstructed analysis regions} \\
Process &    $p_{\textrm{T}}^{V\textrm{, t}}$ interval & Number of leptons & $\pT^{V}$ interval & Number of jets \\
\midrule
$WH$ & 150--250~\GeV          & 1 & 150--250~\GeV & 2, 3 \\
\midrule
$WH$ &  {$>250$~\GeV}                               & 1 & {$>250$~\GeV} & 2, 3 \\
\midrule
$ZH$ & 75--150~\GeV                    &  2 & 75--150~\GeV & 2, $\geq 3$ \\
\midrule
\multirow{2}{*}{$ZH$} & \multirow{2}{*}{{150--250~\GeV}}  &   {0} & {150--250~\GeV}  & {2, 3}  \\
&                              &  {2} & {150--250~\GeV}  & {2, $\geq$3}  \\
\midrule
\multirow{2}{*}{$ZH$} &  \multirow{2}{*}{$>250$~\GeV}              &  {0} & {$>250$~\GeV} & {2, 3} \\
&                             &  {2} & {$>250$~\GeV} & {2, $\geq$3}  \\
\bottomrule
\end{tabular}
\end{center}
\end{table}

\begin{figure}[h]
\begin{center}
\subfloat[]{\includegraphics[width=0.8\linewidth]{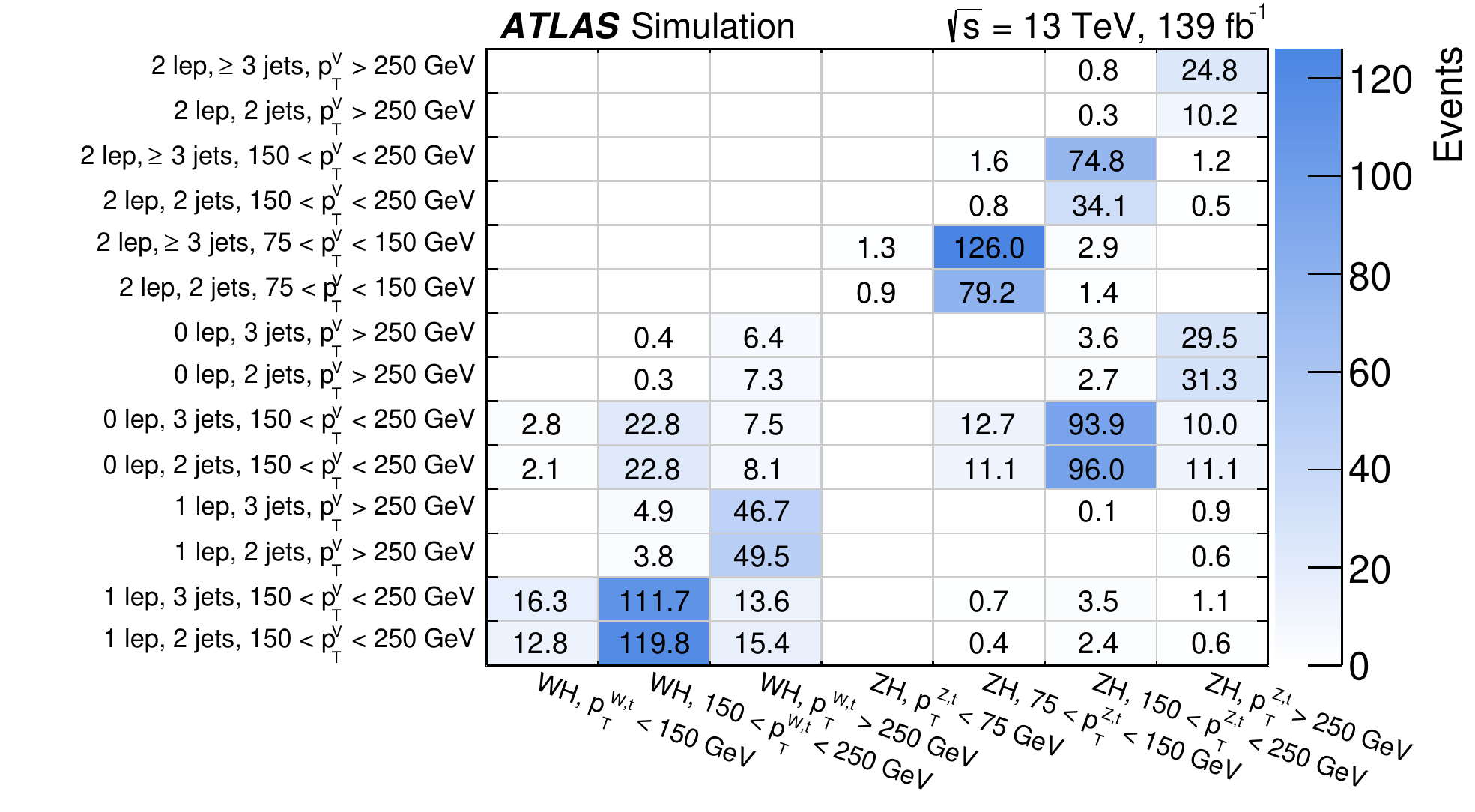}\label{fig:5POI_acceptance}} \\
\subfloat[]{\includegraphics[width=0.8\linewidth]{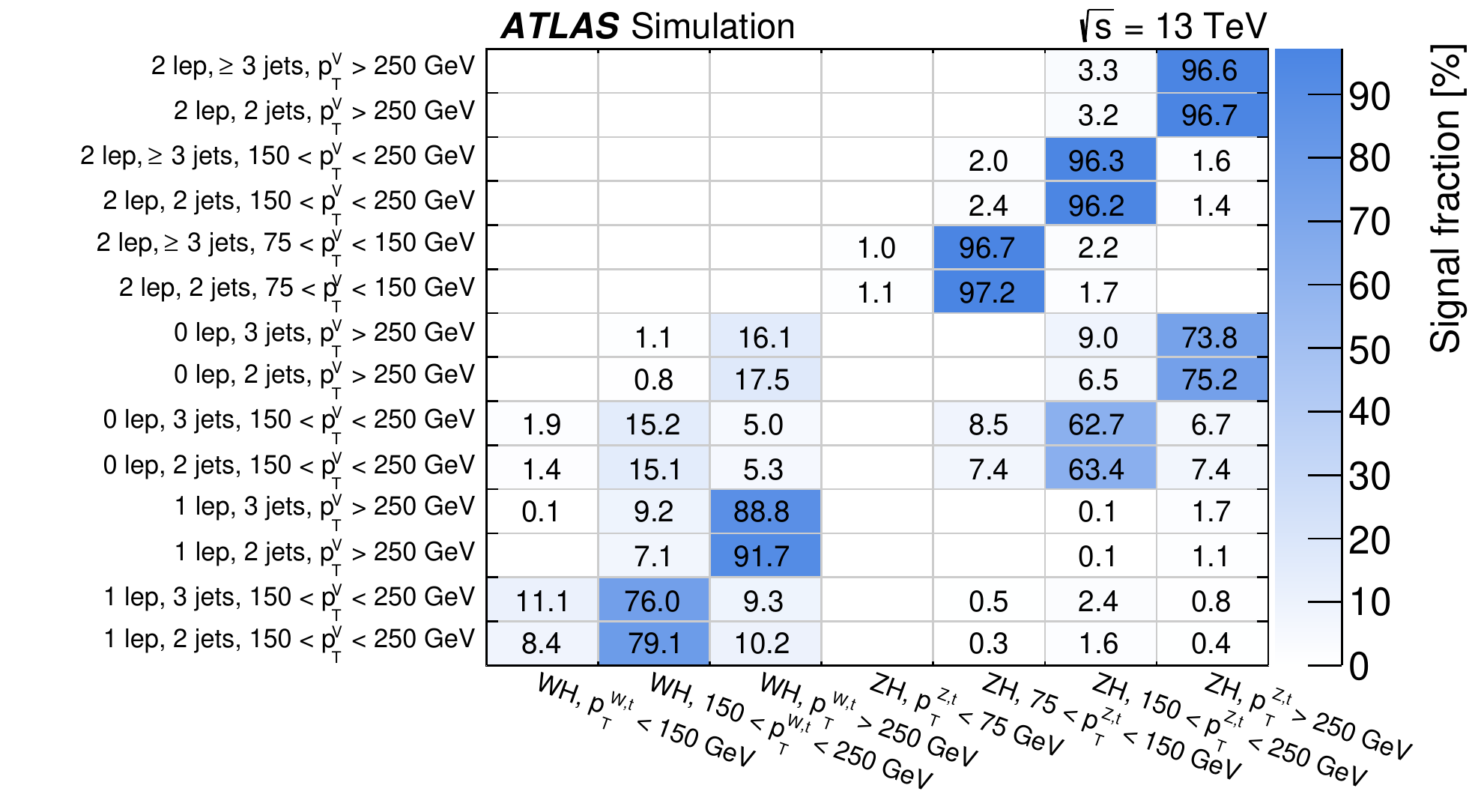}\label{fig:5POI_purity}}
\caption{For each of the STXS regions,
(a) the predicted signal event yield for $VH$, $V\to$~leptons, $H\to b\bar{b}$ events
of each reconstructed-analysis region ($y$-axis) for each STXS signal region ($x$-axis);
(b) the predicted fraction of signal events passing all selection criteria (in percent) in every reconstructed-event category ($y$-axis) from each STXS signal region
($x$-axis).
Entries with event yield below 0.1 or signal
fractions below 0.1\% are not shown.
}
\end{center}
\end{figure}
 
\section{Multivariate discriminants}
A multivariate discriminant is used to improve the sensitivity of the analysis.
Two sets of boosted decision trees (BDTs) are trained using the same input variables. A nominal set, referred to as BDT$_{VH}$, is designed to discriminate the $VH$ signal from the background processes. A second set, referred to as BDT$_{VZ}$, which aims to separate the $VZ, Z\rightarrow b\bar{b}$ diboson process from the $VH$ signal and other background processes, is used to validate the $VH$ analysis.
In each set, BDTs are trained in eight regions, obtained by merging some of the 14 analysis regions. In particular, the $150\,\GeV~<$~\ptv~$<~250\,\GeV$ and \ptv~$>~250\,\GeV$ analysis regions in each lepton channel and jet category are merged for the training, as no increase in sensitivity was found when undertaking separate trainings in the two regions.
The outputs of the BDTs, evaluated in each signal region, are used as final discriminating variables.

The BDT input variables used in the three lepton channels are detailed in Table~\ref{tbl:MVAinputs}. The separation of two $b$-tagged jets in pseudorapidity is denoted by $|\Delta \eta(\vec{b_{1}}, \vec{b_{2}})|$. In 3-jet events, the third jet is labelled
as jet$_3$ and the mass of the 3-jet system is denoted $m_{bbj}$. The azimuthal angle between the vector boson
and the system of the Higgs boson candidate formed from the two $b$-tagged jets is denoted $\Delta \phi(\vec{V}, \vec{bb})$, and their pseudorapidity separation is denoted
$\Delta \eta(\vec{V}, \vec{bb})$. In the 0-lepton channel, $m_{\text{eff}}$ is defined as the scalar sum of the transverse momenta of all
jets and the \met\ ($m_{\text{eff}} = H_\text{T} + \met$). In the 1-lepton channel, the angle between the lepton and the closest
$b$-tagged jet in the transverse plane is denoted $\min(\Delta\phi(\vec{\ell},\vec{b}))$ and two variables are used to improve the rejection of the \ttbar\ background: the rapidity difference between the $W$ and Higgs boson candidates, $|\Delta y(\vec{V},\vec{bb})|$ and, assuming that the event is \ttbar, the reconstructed top quark mass, $m_{\text{top}}$. The latter is calculated as the invariant mass of the lepton, the reconstructed neutrino and the $b$-tagged jet that yields the lower mass value. For both variables, the transverse component of the neutrino momentum is identified with $\vecmet$, and the longitudinal component is obtained by applying a $W$-mass constraint to the lepton--neutrino system. The variable $\met/\sqrt{S_{\mathrm{T}}}$, where $S_{\mathrm{T}}$ is the scalar sum of transverse momenta of the charged leptons and jets in the event, is defined for use in the 2-lepton channel.

In addition to the above, which were all used in the previous iteration of the analysis~\cite{HIGG-2018-04}, the following variables are also input to the BDTs:
\begin{itemize}
\item Binned MV2 $b$-tagging discriminant: The MV2 discriminant for the two $b$-tagged jets is input to the BDT. The MV2 discriminant is grouped into two bins corresponding to efficiencies of 0--60\% and 60\%--70\%, which are calibrated to data~\cite{FTAG-2018-01, ATLAS-CONF-2018-001, ATLAS-CONF-2018-006}. This variable provides additional rejection against backgrounds where a $c$-jet or light-flavour jet has been misidentified as a $b$-jet, especially $W\rightarrow cq$ in the $\ttbar$ and $Wt$ backgrounds.  This improves the sensitivity in the 1-lepton (0-lepton) channel by $\sim10\%$ ($\sim7\%$). The binned MV2 discriminant does not provide any additional sensitivity in the 2-lepton channel, where the backgrounds are dominated by processes containing two $b$-jets.
\item Magnitude of the track-based \met\ soft-term, $p_{\mathrm{T}}^{\mathrm{miss},\mathrm{st}}$:  In the 0-lepton channel this provides additional rejection against the \ttbar\ background, which may contain unreconstructed objects, such as leptons or $b$-jets, due to kinematic and detector acceptance. The presence of such objects in an event will result in a larger $p_{\mathrm{T}}^{\mathrm{miss},\mathrm{st}}$ for \ttbar\ events than for signal events. This improves the sensitivity in the 0-lepton channel by $\sim2\%$--3\%.
\item $Z$ boson polarisation, $\cos{\theta(\vec{\ell^-},\vec{Z})}$: The $\cos{\theta(\vec{\ell^-},\vec{Z})}$ is calculated as the cosine of the polar angle between the lepton ($\ell^-$) direction in the $Z$ rest frame and the flight direction of the $Z$ boson in the laboratory frame. The $Z$ bosons from the $ZH$ signal process are expected to have a different polarisation compared to those from the dominant $Z$+jets background~\cite{Goncalves:2018fvn}, which provides additional background rejection in the 2-lepton channel. This improves the sensitivity in the 2-lepton channel by $\sim7\%$.
\end{itemize}
 
The distributions of all input variables of the BDTs are compared between data and simulation, and good agreement is found within the uncertainties. The same training procedures and BDT output binning transformation as those detailed in Ref.~\cite{HIGG-2018-04} are used, with the exception that the training algorithm was updated to use gradient boosting in the TMVA~\cite{TMVA} framework.
 
\begin{table}[htb!]
\begin{center}
\caption{Variables used for the multivariate discriminant in each of the
channels, where the $\times$ symbol indicates the inclusion of a variable.}
\label{tbl:MVAinputs}
\begin{tabular}{lccc}
\toprule
Variable                                      &  0-lepton   & 1-lepton    & 2-lepton    \\
\midrule
$m_{{bb}}$                                    &  $\times$ & $\times$  & $\times$  \\
$\Delta R(\vec{b_1},\vec{b_2})$                      &  $\times$ & $\times$  & $\times$  \\
$\pt^{{b}_1}$                                 &  $\times$ & $\times$  & $\times$  \\
$\pt^{{b}_2}$                                 &  $\times$ & $\times$  & $\times$  \\
\ptv                                          &     $\equiv \met $       & $\times$  & $\times$  \\
$\Delta\phi(\vec{V},\vec{bb})$                         &  $\times$ & $\times$  & $\times$  \\
MV2($b_{1}$) &  $\times$ & $\times$ &  \\
MV2($b_{2}$) &  $\times$ & $\times$ & \\
$|\Delta\eta(\vec{b_1},\vec{b_2})|$                  &  $\times$ &           &   \\
$m_{\mathrm{eff}}$                               &  $\times$ &             &             \\
$p_{\mathrm{T}}^{\mathrm{miss},\mathrm{st}}$ & $\times$ & & \\
\met                                          &  $\times$ & $\times$  &           \\
$\mathrm{min}[\Delta\phi(\vec{\ell},\vec{b})]$         &             & $\times$  &          \\
\mtw                                          &             & $\times$  &             \\
$|\Delta y(\vec{V},\vec{bb})|$                        &   & $\times$  &   \\
\mtop                        &   & $\times$  &   \\
$|\Delta\eta(\vec{V},\vec{bb})|$                       &             &             & $\times$  \\
$\met/\sqrt{S_{\mathrm{T}}}$                                          &             &             & $\times$  \\
\mll                                          &             &             & $\times$  \\
$\cos{\theta(\vec{\ell^-},\vec{Z})}$ & & & $\times$ \\
\midrule
& \multicolumn{3}{c}{Only in 3-jet events} \\
$\pt^{\mathrm{jet}_3}$                          &  $\times$ & $\times$  & $\times$  \\
$m_{bbj}$                                     &  $\times$ & $\times$  & $\times$  \\
\bottomrule
\end{tabular}
\end{center}
\end{table}
 
\section{Background modelling}
The simulated event samples summarised in Section~\ref{sec:samples} are used to model all background processes, except for the \ttbar\ background in the 2-lepton channel\footnote{The \ttbar\ background in the 2-lepton channel was modelled using simulated event samples in the previous publication~\cite{HIGG-2018-04}.} and the multi-jet background in the 1-lepton channel, which are both estimated using data-driven techniques, as discussed below.
 
\subsection{Data-driven \ttbar background estimation} \label{sec:ttbardatadriven}
 
In the 2-lepton channel a high-purity control region, over 99\% pure in \ttbar and single-top-quark $Wt$ events (jointly referred to as the top background), is defined using the nominal event selection, but replacing the same-flavour lepton selection with a requirement of exactly one electron and one muon. This region is referred to as the $e\mu$-control region, $e\mu$-CR. As these top background events typically contain two $W$ bosons which decay into leptons, they are symmetric in lepton flavour. The events in the $e\mu$-CR are directly used to model the shape and normalisation of the same-flavour lepton top background in the nominal selection. Any bias caused from the lepton trigger, reconstruction, identification or acceptance, is determined by comparing the yield of simulated top background events in the nominal selection with that in the $e\mu$ control region. No significant bias in the shape or normalisation is observed for any of the important kinematic variables, including the BDT discriminant. A ratio of the top yield in the analysis region to that in the $e\mu$-CR of $1.00\pm 0.01$ ($1.01\pm 0.01$) is determined using simulation, for the 2-jet ($\geq 3$-jet) region, where the uncertainty in the ratio is the statistical uncertainty resulting from the simulated samples. As no evaluated theoretical or experimental uncertainties create any bias beyond the statistical uncertainty of the ratio, the latter is assigned as an extrapolation uncertainty. This method has the advantage that all the experimental and theoretical uncertainties are eliminated, resulting in the data statistics in the $e\mu$-CR becoming the dominant uncertainty source for the data-driven top background estimate.

\subsection{Multi-jet background estimation}
 
Multi-jet (MJ) event production has a large cross-section and thus, despite not being a source of genuine missing transverse momentum or prompt leptons, has the potential to contribute a non-negligible amount of background. Using the same techniques detailed in Ref.~\cite{HIGG-2018-04}, the MJ background was demonstrated to be negligible in both the 0- and 2-lepton channels.
 
In the 1-lepton channel, the MJ background is reduced to the percent level and is predicted using the same method as described in Ref.~\cite{HIGG-2018-04} with minor changes to account for the use of the MV2($b_j$) variables in the BDT.
The MJ background is modelled from data in an MJ-enriched control region (MJ-CR), from which all simulated backgrounds are subtracted. The MJ-CR is defined by applying the nominal event selection, except for the stricter lepton isolation requirement, which is inverted. The requirement on the number of $b$-tagged jets is relaxed from two (2-$b$-tag MJ-CR) to one (1-$b$-tag MJ-CR) to increase the statistical precision. To correctly estimate the 2-$b$-tag MJ BDT shape, the values of both the MV2($b_1$) and MV2($b_2$) BDT input variables in the 1-$b$-tag events, are replaced with values emulated from a joint MV2($b_1$) and MV2($b_2$) probability distribution derived from the 2-$b$-tag MJ-CR. The normalisation of the MJ background is then determined from a template fit to the \mtw distribution after applying the nominal selection with a 2-$b$-tag requirement, using the MJ shape predicted  from the 1-$b$-tag MJ-CR and the shapes of the other backgrounds from simulation.

 
\section{Systematic uncertainties}
\label{sec:syst}
The sources of systematic uncertainty can be broadly divided into three groups: those of an experimental nature, those related
to the modelling of the backgrounds and those associated with the Higgs boson signal simulation. The estimation of the uncertainties closely follows the methodology outlined in Refs.~\autocite{HIGG-2016-29, HIGG-2018-50} and is briefly summarised below.
 
\subsection{Experimental uncertainties}
 
The dominant experimental uncertainties originate from the $b$-tagging correction factors, jet energy scale calibration and the modelling of the jet energy resolution. The $b$-tagging correction factors, determined from the difference between the efficiencies measured in data and simulation, are evaluated in five MV2 discriminant bins and are derived separately for $b$-jets, $c$-jets and light-flavour jets~\autocite{FTAG-2018-01, ATLAS-CONF-2018-001, ATLAS-CONF-2018-006}.  All of the correction factors for the three jet flavours have uncertainties estimated from multiple measurements, which are decomposed into uncorrelated components that are then treated independently. 
The uncertainties in the jet energy scale and resolution are based on their respective measurements~\autocite{PERF-2016-04,PERF-2011-04}. 
 
Uncertainties in the reconstruction, identification, isolation and trigger efficiencies of muons~\autocite{PERF-2015-10}
and electrons~\cite{PERF-2017-01} are considered, along with the uncertainty in their energy scale and resolution. These are found to have only a small impact on the result.
The uncertainties in the energy scale and resolution of the jets and leptons are propagated to the calculation of \met, which also has additional uncertainties from the modelling
of the underlying event and momentum scale, momentum resolution
and reconstruction efficiency of the tracks used to compute the soft-term~\autocite{PERF-2016-07, ATLAS-CONF-2018-023}. An uncertainty is assigned to the \met\ trigger correction factors, determined from the ratio of the trigger efficiency in data and simulation, to account for the statistical uncertainty in the measured correction factors and for differences between the correction factors determined from $W$~+~jets, $Z$~+~jets and $t\bar{t}$ events.
The uncertainty in the combined 2015--2018 integrated luminosity is 1.7\%. It is derived following a methodology similar to that detailed in Ref.~\cite{DAPR-2013-01}, and using
the LUCID-2 detector for the baseline luminosity measurements~\cite{Avoni:2018iuv}.
The average number of interactions per bunch crossing in the simulation is rescaled by 1.03 to improve agreement between simulation and data, based on the measurement of the visible cross-section in minimum-bias events~\autocite{STDM-2015-05}, and an uncertainty, as large as the correction, is included.
 
\subsection{Background uncertainties}
 
Modelling uncertainties are derived for the simulated samples and broadly cover three areas: normalisations (referred to as normalisation uncertainties), acceptance
differences that affect the relative normalisations between regions with a common underlying normalisation (referred to as relative acceptance uncertainties), and the
shapes of the differential distributions of the kinematic variables (referred to as shape uncertainties).
 
The overall cross-sections and associated
normalisation uncertainties for the background processes are taken from the currently most accurate calculations as detailed in
Table~\ref{tab:samples}, apart from the main backgrounds (\zhf, \whf, \ttb) whose normalisations are left unconstrained (floated) in the global
likelihood fit.
 
The relative acceptance and shape uncertainties are derived from either
particle-level or reconstruction-level comparisons between nominal and alternative simulated samples, or from comparisons with data in control regions.
The alternative samples are produced either by different generators or by altering the nominal generator's parameter values.
When relative acceptance uncertainties are estimated, the nominal and alternative samples are normalised
using the same production cross-section.
Shape uncertainties are estimated within a signal region, an analysis region or a set of analysis regions, depending on the distribution being varied, with the nominal and alternative samples scaled to have the same normalisation in the studied area.
Shape uncertainties over regions with different acceptance, can affect not only the shape, but also cause event migration between regions (referred to as a shape plus migration uncertainty) as opposed to an uncertainty that only alters the shape within a single SR (referred to as just a shape uncertainty).
Unless stated otherwise, the uncertainty is taken from the alternative sample that differs most in shape from the nominal sample.

Shape uncertainties for \zhf, single-top and diboson backgrounds are derived for the \mbb\ and \ptv\ variables, as it was found sufficient to consider the changes induced in these variables to cover the overall shape variation of the BDT discriminant. For \whf\ and \ttb\ backgrounds, a more sophisticated multidimensional parameterisation method is introduced to estimate the shape uncertainties of the final discriminant~\cite{Cranmer:2015bka,}. In this method, a BDT (referred to as BDT$_S$) is trained to discriminate the nominal sample from an alternative sample, using the kinematic variables from the BDT$_{VH}$ (Table~\ref{tbl:MVAinputs}) as input variables, except for the \ptv. Before training, the \ptv\ distribution of the nominal sample is reweighted to match that of the alternative sample. The \ptv\ difference is considered as a separate, uncorrelated uncertainty, in a manner similar to that for the other backgrounds. The ratio of the BDT$_S$ distributions evaluated for the alternative and nominal samples provide a reweighting function (referred to as R$_{\textrm{BDT}}$), which can be used to correct the nominal sample to match the alternative sample. This method simultaneously maps the $n$-dimensional space formed by the kinematic variables
of the two generators onto each other. It is verified that, after being reweighted by R$_{\textrm{BDT}}$, the input variable distributions for the nominal sample are in good agreement with those of the alternative sample.
 
The systematic uncertainties affecting the modelling of the background samples are summarised in
Tables~\ref{tab:bkg_systematics} and~\ref{tab:bkgdib_systematics}, and key details of the treatment of the backgrounds are
reported below.
 
\begin{table}[p!]
\begin{small}
\begin{center}
\caption
{Summary of the systematic uncertainties in the background modelling for
$Z$~+~jets, $W$~+~jets, \ttb, single-top-quark and multi-jet production. `ME' indicates a matrix element generator variation and `PS' indicates a parton shower generator variation.
An `M+S' symbol is used when a shape uncertainty includes a migration effect that allows relative acceptance changes between regions, whilst `S' indicates that the uncertainty only acts upon the shape in the signal region.
Instances where an uncertainty is considered independently in different regions are detailed in parentheses. Where the size of an acceptance systematic uncertainty varies between regions, a range is displayed.
\label{tab:bkg_systematics}}
\begin{tabular}{ l  c}
\toprule
\multicolumn{2}{c}{$Z$~+~jets}\\
\midrule
$Z+ll$ normalisation & 18\% \\
$Z+cl$ normalisation & 23\% \\
\multirow{2}{*}{$Z$~+~HF normalisation} & Floating (2-jet, 3-jet) $\times$   \\
& ($75\,\GeV<\ptv<150\,\GeV, \ptv > 150\,\GeV$) \T \\
$Z+bc$-to-$Z+bb$ ratio & 30\%--40\% \\
$Z+cc$-to-$Z+bb$ ratio & 13\%--16\% \\
$Z+bl$-to-$Z+bb$ ratio & 20\%--28\% \\
SR-to-low $\Delta$R CR ratio & 3.8\%--9.9\% ($75\,\GeV<\ptv<150\,\GeV, \ptv > 150\,\GeV$) \T \\
SR-to-high $\Delta$R CR & 2.7\%--4.1\%  ($75\,\GeV<\ptv<150\,\GeV, \ptv > 150\,\GeV$) \T \\
0-to-2 lepton ratio & 7\% \\
\ptv  & M+S ($75\,\GeV<\ptv<150\,\GeV, \ptv > 150\,\GeV$)  \\
\mbb  & S ($75\,\GeV<\ptv<150\,\GeV, \ptv > 150\,\GeV$) \\
\midrule
\multicolumn{2}{c}{$W$~+~jets}\\
\midrule
$W+ll$ normalisation & 32\% \\
$W+cl$ normalisation & 37\% \\
$W$~+~HF normalisation & Floating (2-jet, 3-jet)\\
$W+bc$-to-$W+bb$ ratio & 15\% (0-lepton) and 30\% (1-lepton)    \\
$W+cc$-to-$W+bb$ ratio & 10\% (0-lepton) and 30\% (1-lepton) \\
$W+bl$-to-$W+bb$ ratio & 26\% (0-lepton) and 23\% (1-lepton) \\
SR-to-CR ratio & 3.6\%--15\% \\
0-to-1 lepton ratio & 5\%  \\
\ptv & M+S (2-jet, 3-jet) \\
R$_{\textrm{BDT}}$ & S \\
\midrule
\multicolumn{2}{c}{\ttb\ (0+1-lepton channels only)}\\
\midrule
\ttb\ normalisation & Floating (2-jet, 3-jet) \\
0-to-1 lepton ratio & 8\% \\
\ttbar (flavour composition) $bc$-to-$bb$ ratio (ME) &  7.6\%--8.2\% (0-lepton), 1.3\%--3.8\% (1-lepton) \\
\ttbar (flavour composition) $bc$-to-$bb$ ratio (PS) &  2.1\%--3.2\% (0-lepton), 1.5\%--7.1\% (1-lepton) \\
\ttbar (flavour composition) other-to-$bb$ ratio (ME) &   2.8\%--6.4\% (0-lepton), 3.3\%--5.7\% (1-lepton) \\
\ttbar (flavour composition) other-to-$bb$ ratio (PS) &  5.6\%--13\% (0-lepton), 0.3\%--2.1\% (1-lepton) \\
\ptv & M+S (2-jet, 3-jet) \\
R$_{\textrm{BDT}}$ ME variation & M+S (2-jet, 3-jet) \\
R$_{\textrm{BDT}}$ PS variation  & M+S (0-lepton, 1-lepton) \\
\midrule
\multicolumn{2}{c}{Single top quark}\\
\midrule
Cross-section & 4.6\% ($s$-channel), 4.4\% ($t$-channel), 6.2\% ($Wt$) \\
Acceptance 2-jet & 17\% ($t$-channel), 55\% ($Wt(bb)$), 24\% ($Wt(\mathrm{other})$) \\
Acceptance 3-jet & 20\% ($t$-channel), 51\% ($Wt(bb)$), 21\% ($Wt(\mathrm{other})$)  \\
\mbb & M+S ($t$-channel, $Wt(bb)$, $Wt(\mathrm{other})$) \\
\ptv & M+S ($t$-channel, $Wt(bb)$, $Wt(\mathrm{other})$) \\
\midrule
\multicolumn{2}{c}{Multi-jet (1-lepton)}\\
\midrule
Normalisation & 30\%--200\% (2-jet), 100\% (3-jet) \\
BDT template & M+S \\
\bottomrule
\end{tabular}
\end{center}
\end{small}
\end{table}

\begin{table}[ht!]
\begin{small}
\begin{center}
\caption
{Summary of the systematic uncertainties in the background modelling for diboson production. `PS/UE' indicates parton shower/underlying event. An `M+S' symbol is used when a shape uncertainty includes a migration effect that allows relative acceptance changes between regions. Instances where an uncertainty is considered independently in different regions are detailed in parentheses.
When extracting the $(W/Z)Z$ diboson production signal yield, as the normalisations are unconstrained,
the normalisation uncertainties are removed. Where the size of an acceptance systematic uncertainty varies between regions, a range is displayed.
\label{tab:bkgdib_systematics}}
\begin{tabular}{ l  c}
\toprule
\multicolumn{2}{c}{$ZZ$}\\
\midrule
Normalisation & 20\% \\
0-to-2 lepton ratio & 6\% \\
Acceptance from scale variations& 10\%--18\% \\
Acceptance from PS/UE variations for 2 or more jets & 6\% \\
Acceptance from PS/UE variations for  3  jets & 7\% (0-lepton), 3\% (2-lepton)\\
\mbb\ from  scale variations& M+S (correlated with $WZ$ uncertainties)\\
\ptv\ from  scale variations& M+S (correlated with $WZ$ uncertainties)\\
\mbb\ from PS/UE variations& M+S (correlated with $WZ$ uncertainties)\\
\ptv\ from PS/UE variations& M+S (correlated with $WZ$ uncertainties)\\
\mbb\  from matrix-element variations& M+S (correlated with $WZ$ uncertainties)\\
\midrule
\multicolumn{2}{c}{$WZ$}\\
\midrule
Normalisation & 26\% \\
0-to-1 lepton ratio & 11\% \\
Acceptance from scale variations & 13\%--21\%   \\
Acceptance from PS/UE variations for 2 or more jets & 4\%\\
Acceptance from PS/UE variations for  3  jets & 11\% \\
\mbb\  from  scale variations& M+S (correlated with $ZZ$ uncertainties)\\
\ptv\  from  scale variations& M+S (correlated with $ZZ$ uncertainties)\\
\mbb\ from PS/UE variations& M+S (correlated with $ZZ$ uncertainties)\\
\ptv\ from PS/UE variations& M+S (correlated with $ZZ$ uncertainties)\\
\mbb\ from matrix-element variations& M+S (correlated with $ZZ$ uncertainties)\\
\midrule
\multicolumn{2}{c}{$WW$} \\
\midrule
Normalisation & 25\% \\
\bottomrule
\end{tabular}
\end{center}
\end{small}
\end{table}

\textbf{\vjets\ production} The \vjets\ backgrounds are subdivided into three different components based upon the jet flavour
labels of the two $b$-tagged jets in the event. The main background contributions ($V+bb$, $V+bc$, $V+bl$ and $V+cc$)
are jointly considered as the \vhf\ background. Their overall normalisations are free
to float in the global likelihood fit, separately in the 2- and 3-jet categories. For the \zhf\ background, the normalisations are also floated separately in
the 75~\GeV~$<$~\ptv~$<~150$~\GeV\ and \ptv~$>~150$~\GeV\ regions. The remaining flavour components, \vcl\ and \vll, constitute less than $\sim1$\% of the
background in each analysis region and only normalisation uncertainties are included.
 
Uncertainties are estimated for the relative normalisation of the four heavy-flavour components that constitute the \vhf\ background. These are taken as uncertainties in the $bc$, $cc$ and $bl$ yields compared with the dominant $bb$ yield and are estimated separately in each lepton channel in a manner similar to the acceptance systematic uncertainties.
Relative acceptance uncertainties for the \whf\ background are estimated for the ratio of the event yield in the 0-lepton channel to that in the 1-lepton channel. For the \zhf\ background, there is a relative acceptance uncertainty in the ratio of the event yield in the 0-lepton channel to that in the 2-lepton channel in the \ptv~$> 150$~\GeV\ region. For both \whf\ and \zhf, relative acceptance uncertainties are estimated for the ratio of the event yield in the SR to that in the CRs.
 
For \zhf, shape uncertainties are derived for \mbb\ and \ptv, which are evaluated from comparisons with data in the \mbb\ side-bands (\mbb~$< 80$~GeV or \mbb $> 140$~\GeV), after subtracting backgrounds other than $Z$~+~jets. For \whf, uncertainties are derived for \ptv\ and the R$_{\textrm{BDT}}$ method from comparisons of the nominal sample (\textsc{Sherpa}) with an alternative sample (\textsc{MadGraph}5\_aMC@NLO+\PYTHIAV{8}~\cite{Alwall:2011uj, Bahr:2008pv}).
 
\textbf{\ttb\ production}  In the 0- and 1-lepton channels (jointly referred to as 0+1-lepton channel) separate floating normalisations are used for the 2-jet region and 3-jet region. Uncertainties are derived from comparisons between the nominal sample (\textsc{Powheg}+\PYTHIAV{8}) and alternative samples corresponding to matrix-element (\textsc{MadGraph}5\_aMC@NLO+\PYTHIAV{8}) and parton-shower (\textsc{Powheg}+\textsc{Herwig} 7~\cite{Bellm:2015jjp}) generator 	variations.
 
Relative acceptance uncertainties are estimated for the 0-lepton and 1-lepton channel normalisation ratios. The dominant flavour component of the two $b$-tagged jets in \ttb\ is $bb$. However, there is a sizeable $bc$ component which has a more signal-like topology. Uncertainties in the relative composition of three components, $bb$, $bc$, and any other flavour configuration (referred to as `other') are estimated from the difference in the ratio of the $bc$ or other components to the $bb$ yield between the nominal sample and the alternative matrix element and parton shower generator samples. Shape uncertainties are derived for \ptv\ and using the R$_{\textrm{BDT}}$ method in the 0+1-lepton channels from comparisons with the alternative parton shower and matrix element generator samples.
 
In the 2-lepton channel the \ttb\ background is estimated by a data-driven method as discussed in Section~\ref{sec:ttbardatadriven}. The uncertainty in this background is dominated by the statistical uncertainty of the $e\mu$ control region data events.
 
\textbf{Single-top-quark production} In the $Wt$- and $t$-channels, uncertainties are derived for the normalisation,
relative acceptance and shapes of the \mbb\ and \ptv\ distributions. For the $Wt$-channel, the estimated modelling uncertainties are applied independently according to the flavour of the two $b$-tagged jets, due to the different regions of phase space being probed when there are two $b$-jets ($bb$) present compared with events where there are fewer $b$-jets present (referred to as `other'). Those uncertainties are evaluated from comparisons between the nominal sample (\textsc{Powheg}+\PYTHIAV{8} using the diagram removal scheme~\cite{Frixione:2008yi}) and alternative samples with parton-shower variations (\textsc{Powheg}+\textsc{Herwig}++) and a different scheme to account for the interference between $Wt$ and \ttb\ production (\textsc{Powheg}+\PYTHIAV{8} using the diagram subtraction scheme)~\cite{Frixione:2007vw}.
Only a normalisation uncertainty is derived for the $s$-channel, since its contribution is at a very low level.
 
\textbf{Diboson production} The diboson backgrounds are composed of three distinct processes: $WZ$, $WW$ and $ZZ$
production. Given the small contribution from $WW$ production ($<0.1\%$ of the total background) only a normalisation
uncertainty is assigned. For the more important contributions from the $WZ$ and $ZZ$ backgrounds, uncertainties are considered in
the overall normalisation, the relative acceptance between regions and the \mbb\ and \ptv\ shapes. These are derived following the procedure described in Ref.~\autocite{HIGG-2016-29} and are outlined in Table~\ref{tab:bkgdib_systematics}, which includes comparisons of the nominal sample (\textsc{Sherpa}) with alternative samples (\textsc{Powheg}+\PYTHIAV{8} and \textsc{Powheg}+\textsc{Herwig}++).
 
\textbf{Multi-jet background uncertainties} The systematic uncertainties in the multi-jet background estimate in the 1-lepton channel are derived by following the procedure outlined in Ref.~\cite{HIGG-2018-04}. Two different uncertainty components are considered, those which alter the normalisation and those which alter the multi-jet BDT template shape.
 
\subsection{Signal uncertainties}
 
The systematic uncertainties that affect the modelling of the signal are summarised in Table~\ref{tab:sig_systematics}  and are estimated with procedures that closely follow those outlined in Ref.~\cite{HIGG-2018-50, ATL-PHYS-PUB-2018-035, Bendavid:2018nar, deFlorian:2016spz}. The systematic uncertainties in the calculations of the $VH$ production cross-sections and the $H\to\bbbar$ branching fraction\footnote{These systematic uncertainties are fully degenerate with the signal yield and do not affect the calculation of the significance relative to the background-only prediction and STXS cross-section measurement.} are assigned following the recommendations of the LHC Higgs Cross Section Working Group~\cite{Dittmaier:2011ti,Dittmaier:2012vm,Heinemeyer:2013tqa,Harlander:2013mla,Brein:2012ne}.
 
Uncertainties in the \mbb\ and \ptv\ signal shape are estimated, as described in Ref.~\cite{HIGG-2018-04}, from scale variations, PDF and $\alphas$ (PDF+$\alphas$) uncertainties, from varying the parton shower and underlying event (PS/UE) models using AZNLO tuning variations and from comparisons with alternative parton-shower generator samples (\textsc{Powheg}+\textsc{Herwig} 7). In addition, a systematic uncertainty from higher-order EW corrections effects is taken into account as a variation in the shape of the \ptv distributions for $qq\rightarrow VH$ production. 
Acceptance uncertainties, evaluated according to STXS regions, correctly accounting for the migration and correlations between regions, are evaluated for the scale variations, PS/UE models and PDF+$\alphas$.
 
For the STXS measurement, the signal uncertainties are separated into two groups,
uncertainties in the acceptance and shape of kinematic distributions which alter the signal modelling (theoretical modelling uncertainties) and the uncertainties in the prediction of the production cross-section for each of these regions
(theoretical cross-section uncertainties). Whilst theoretical modelling uncertainties enter the STXS measurements, theoretical cross-section uncertainties only affect the predictions with which they are compared, and are therefore not included in the likelihood function.
 
\begin{table}[ht!]
\begin{small}
\begin{center}
\caption
{Summary of the systematic uncertainties in the signal modelling. `PS/UE' indicates parton shower/underlying event. An `M+S' symbol is used when a shape uncertainty includes a migration effect that allows relative acceptance changes between regions. Instances where an uncertainty is considered independently in different regions are detailed in parenthesis. Where the size of an acceptance systematic uncertainty varies between regions, a range is displayed.
\label{tab:sig_systematics}}
\begin{tabular}{ l  c}
\toprule
\multicolumn{2}{c}{Signal}\\
\midrule
Cross-section (scale) & 0.7\% ($qq$), 25\% ($gg$) \\
$H \to \bbbar$ branching fraction & 1.7\% \\
Scale variations in STXS bins  & 3.0\%--3.9\% ($qq\rightarrow WH$), 6.7\%--12\% ($qq\rightarrow ZH$), 37\%--100\% ($gg\rightarrow ZH$)  \\
PS/UE variations in STXS bins & 1\%--5\% for $qq\rightarrow VH$, 5\%--20\% for $gg\rightarrow ZH$ \\
PDF+$\alphas$ variations in STXS bins & 1.8\%--2.2\% ($qq\rightarrow WH$), 1.4\%--1.7\% ($qq\rightarrow ZH$), 2.9\%--3.3\% ($gg\rightarrow ZH$)\\
\mbb\ from scale variations& M+S ($qq\rightarrow VH$, $gg\rightarrow ZH$) \\
\mbb\ from PS/UE variations& M+S \\
\mbb\ from PDF+$\alphas$ variations & M+S \\
\ptv\ from NLO EW correction & M+S \\
\bottomrule
\end{tabular}
\end{center}
\end{small}
\end{table}
 
 
\section{Statistical analysis}
 
The statistical procedure is based on a likelihood function $\mathcal{L}(\mu,
\vec{\theta})$, constructed as the product of Poisson probability terms over the bins
of the input distributions, with parameters of interest (POI) extracted by maximising the likelihood. The effects of systematic uncertainties enter the likelihood as nuisance parameters (NP),
$\vec{\theta}$. Most of the uncertainties discussed in
Section~\ref{sec:syst} are constrained with Gaussian or log-normal probability density
functions. The normalisations of the largest backgrounds, $t\bar{t}$, \whf\ and \zhf, can be reliably determined by the
fit, so they are left unconstrained in the likelihood. The uncertainties due to the
limited number of events in the simulated samples used for the background predictions are
included using the Beeston--Barlow technique~\autocite{Barlow:1993dm}. As detailed in
Ref.~\cite{HIGG-2013-23}, systematic variations that are subject to large statistical
fluctuations are smoothed, and systematic uncertainties that have a negligible impact on
the final results are pruned away region-by-region (treating signal and control regions separately).

The global likelihood fit comprises 14 signal
regions, defined as the 2- and 3-jet categories in the two high-\ptv\ ($150<\ptv<250~$\GeV\ and {\ptv $>250$~\GeV}) regions for the three
channels, and in the medium-\ptv\ region ($75<\ptv<150~$\GeV) for the 2-lepton channel. The 28 control regions are also input as event yields in all fit configurations.
 
Three different versions of the analysis are studied, which differ in the distributions input to the fit.
\begin{itemize}
\item The nominal analysis, referred to as the multivariate analysis, uses the BDT$_{VH}$
multivariate discriminant output distributions as the inputs to the fit. Three different POI configurations are studied. Firstly, a single-POI fit measures $\mu_{VH}^{bb}$, the signal strength that
multiplies the SM Higgs boson $VH$ production cross-section times the branching fraction into
$\bbbar$. Secondly, a two-POI fit is undertaken, which jointly measures the signal strengths of the $WH$ and $ZH$ components. Finally, a five-POI fit version measures the signal cross-section multiplied by the $H\to b\bar{b}$ and $V\to$~leptons branching fractions in the five STXS regions (see Table~\ref{tab:STXS_reducedstage1}).
\item The dijet-mass cross-check analysis uses the \mbb\ distributions, instead of the BDT$_{VH}$ distributions, as inputs to a single-POI fit to measure $\mu_{VH}^{bb}$.
\item The diboson validation analysis, a measurement of the signal strength of the $WZ$ and $ZZ$ processes, uses the BDT$_{VZ}$ output distributions. The SM Higgs boson is included as a background process normalised to the predicted SM cross-section with an uncertainty of 50\%, which conservatively encompasses the previous measurement and uncertainty~\cite{HIGG-2018-04}. Two POI configurations are evaluated, firstly a single-POI fit to measure $\mu^{bb}_{VZ}$, the signal strength of the combined $WZ$ and $ZZ$ diboson processes, and secondly a two-POI fit to simultaneously measure the  $WZ$ and $ZZ$ signal strengths.
\end{itemize}
The background predictions in all post-fit distributions and tables are obtained by normalising the backgrounds and setting the nuisance parameters according to the values determined by the fit used to measure $\mu_{VH}^{bb}$.

 
\section{Results}
\subsection{Signal strength measurements}
 
The post-fit normalisation factors of the unconstrained backgrounds in the global likelihood fit are shown for the single-POI multivariate analysis in Table~\ref{tab:resultscalefactorsMVA}, the post-fit signal and background yields are shown in Tables~\ref{tbl:yieldTable_01l} and~\ref{tbl:yieldTable_2l}, and Figure~\ref{fig:mva_output} shows the BDT$_{VH}$ output distributions in the high-\ptv\ 2-jet SRs, which are most sensitive to the signal.
 
\begin{table}[hptb]
\begin{center}
\caption{Factors applied to
the nominal normalisations of the \ttb, \whf\, and \zhf\ backgrounds, as obtained from the global likelihood fit to the 13~\TeV\ data for the nominal multivariate analysis.
The errors represent the combined statistical and systematic uncertainties.
\label{tab:resultscalefactorsMVA}}
\begin{tabular}{lc}
\toprule
Process and category & Normalisation factor \\
\midrule
\ttb\ 2-jet & $0.98 \pm 0.09$ \\
\ttb\ 3-jet & $0.93 \pm 0.06$ \\
\whf\ 2-jet &  $1.06 \pm 0.11$ \\
\whf\ 3-jet &  $1.15 \pm 0.09$ \\
\zhf\ 2-jet, $75<\ptv<150~$\GeV\  & $1.28 \pm 0.08$ \\
\zhf\ 3-jet, $75<\ptv<150~$\GeV\  & $1.17 \pm 0.05$ \\
\zhf\ 2-jet, $150$~\GeV~$<\ptv$ &  $1.16 \pm 0.07$ \\
\zhf\ 3-jet, $150$~\GeV~$<\ptv$ &  $1.09 \pm 0.04$ \\
\bottomrule
\end{tabular}
\end{center}
\end{table}
 
 
\begin{sidewaystable}[phtb]
\centering
\small
\sisetup{round-mode=places, round-precision=0, group-digits=integer, group-minimum-digits=6}
\caption{The Higgs boson signal, background and data yields for each signal region in the 0- and 1-lepton channels after the full selection.
The signal and background yields are normalised to the results of the global likelihood fit.
All systematic uncertainties are included in the indicated uncertainties.
An entry of ``--'' indicates that a specific background component is negligible in a certain region, or that no simulated events are left after the analysis selection. \label{tbl:yieldTable_01l}}
\begin{tabular}{l
S[table-alignment=right,table-format=5.0]@{\,$\pm$\,}S[table-alignment=left,table-format=2.0]
S[table-alignment=right,table-format=5.0]@{\,$\pm$\,}S[table-alignment=left,table-format=3.0]
S[table-alignment=right,table-format=5.0]@{\,$\pm$\,}S[table-alignment=left,table-format=2.0]
S[table-alignment=right,table-format=5.0]@{\,$\pm$\,}S[table-alignment=left,table-format=2.0]
S[table-alignment=right,table-format=5.0]@{\,$\pm$\,}S[table-alignment=left,table-format=3.0]
S[table-alignment=right,table-format=5.0]@{\,$\pm$\,}S[table-alignment=left,table-format=3.0]
S[table-alignment=right,table-format=5.0]@{\,$\pm$\,}S[table-alignment=left,table-format=2.0]
S[table-alignment=right,table-format=5.0]@{\,$\pm$\,}S[table-alignment=left,table-format=2.0]
}
\toprule
\multirow{2}{*}{Signal regions} & \multicolumn{4}{c}{0-lepton} & \multicolumn{4}{c}{0-lepton} & \multicolumn{4}{c}{1-lepton} & \multicolumn{4}{c}{1-lepton}\\
\cline{2-17}
& \multicolumn{4}{c}{$150\,\GeV<\ptv<250\,\GeV$} & \multicolumn{4}{c}{$\ptv>250\,\GeV$} & \multicolumn{4}{c}{$150\,\GeV<\ptv<250\,\GeV$} & \multicolumn{4}{c}{$\ptv>250\,\GeV$} \T \B \\ \midrule
Sample  & \multicolumn{2}{c}{2-jet} & \multicolumn{2}{c}{3-jet} & \multicolumn{2}{c}{2-jet} & \multicolumn{2}{c}{3-jet} & \multicolumn{2}{c}{2-jet} & \multicolumn{2}{c}{3-jet} & \multicolumn{2}{c}{2-jet} & \multicolumn{2}{c}{3-jet} \T \B \\ \midrule
$Z$~+~jets & 2846 & 80 & 3830 & 160 & 338 & 13 & 533 & 23 & 102.1 & 5.4 & 207 & 11 & 10.34 & 0.88 & 24.5 & 1.6\\
$W$~+~jets & 634 & 63 & 1500 & 110 & 83.1 & 9.3 & 220 & 19 & 1850 & 160 & 4080 & 270 & 353 & 33 & 935 & 69\\
Single top & 237 & 35 & 770 & 130 & 9.1 & 2.1 & 36.3 & 8 & 990 & 160 & 3570 & 600 & 70 & 15 & 313 & 66\\
$\ttbar$ & 1157 & 76 & 5470 & 220 & 38.6 & 4.8 & 151 & 16 & 4600 & 210 & 21030 & 620 & 188 & 19 & 970 & 58\\
Diboson & 360 & 55 & 333 & 79 & 86 & 13 & 70 & 17 & 229 & 57 & 264 & 83 & 57 & 15 & 68 & 22\\
Multi-jet & \multicolumn{2}{c}{--} & \multicolumn{2}{c}{--} & \multicolumn{2}{c}{--} & \multicolumn{2}{c}{--} & 130 & 47 & 24 & 14 & 16 & 10 & 7.4 & 8.3\\
\midrule
Total bkg. & 5234 & 63 & 11910 & 100 & 554 & 15 & 1011 & 21 & 7899 & 82 & 29170 & 170 & 694 & 23 & 2318 & 42\\
\midrule
\midrule
Signal ($\mu=1.02$) & 147 & 24 & 130 & 22 & 39.9 & 6.3 & 33 & 5.5 & 148 & 24 & 125 & 21 & 52.2 & 8.6 & 44.1 & 7.4\\
\midrule
Data
& \multicolumn{1}{c@{\,\hphantom{$\pm$}\,}}{\tablenum[table-alignment=right,table-format=5.0]{5397}}
&& \multicolumn{1}{c@{\,\hphantom{$\pm$}\,}}{\tablenum[table-alignment=right,table-format=5.0]{11875}}
&& \multicolumn{1}{c@{\,\hphantom{$\pm$}\,}}{\tablenum[table-alignment=right,table-format=5.0]{578}}
&& \multicolumn{1}{c@{\,\hphantom{$\pm$}\,}}{\tablenum[table-alignment=right,table-format=5.0]{1046}}
&& \multicolumn{1}{c@{\,\hphantom{$\pm$}\,}}{\tablenum[table-alignment=right,table-format=5.0]{8044}}
&& \multicolumn{1}{c@{\,\hphantom{$\pm$}\,}}{\tablenum[table-alignment=right,table-format=5.0]{29316}}
&& \multicolumn{1}{c@{\,\hphantom{$\pm$}\,}}{\tablenum[table-alignment=right,table-format=5.0]{727}}
&& \multicolumn{1}{c@{\,\hphantom{$\pm$}\,}}{\tablenum[table-alignment=right,table-format=5.0]{2378}}
\\ \bottomrule
\end{tabular}
\end{sidewaystable}
 
\begin{sidewaystable}[phtb]
\centering
\small
\sisetup{round-mode=places, round-precision=0, group-digits=integer, group-minimum-digits=6}
\caption{The Higgs boson signal, background and data yields for each signal region in the 2-lepton channel after the full selection.
The signal and background yields are normalised to the results of the global likelihood fit. The top background is derived from $e\mu$-CR data.
All systematic uncertainties are included in the indicated uncertainties.
An entry of ``--'' indicates that a specific background component is negligible in a certain region, or that no simulated events are left after the analysis selection. \label{tbl:yieldTable_2l}}
\begin{tabular}{l
S[table-alignment=right,table-format=5.0]@{\,$\pm$\,}S[table-alignment=left,table-format=3.0]
S[table-alignment=right,table-format=5.0]@{\,$\pm$\,}S[table-alignment=left,table-format=3.0]
S[table-alignment=right,table-format=5.0]@{\,$\pm$\,}S[table-alignment=left,table-format=2.0]
S[table-alignment=right,table-format=5.0]@{\,$\pm$\,}S[table-alignment=left,table-format=2.0]
S[table-alignment=right,table-format=5.0]@{\,$\pm$\,}S[table-alignment=left,table-format=1.0]
S[table-alignment=right,table-format=5.0]@{\,$\pm$\,}S[table-alignment=left,table-format=2.0]
}
\toprule
\multirow{2}{*}{Signal regions} & \multicolumn{4}{c}{2-lepton} & \multicolumn{4}{c}{2-lepton} & \multicolumn{4}{c}{2-lepton}\\
\cline{2-13}
& \multicolumn{4}{c}{$75\,\GeV<\ptv<150\,\GeV$} & \multicolumn{4}{c}{$150\,\GeV<\ptv<250\,\GeV$} & \multicolumn{4}{c}{$\ptv>250\,\GeV$}  \T \B \\ \midrule
Sample  & \multicolumn{2}{c}{2-jet} & \multicolumn{2}{c}{$\geq$3-jet} & \multicolumn{2}{c}{2-jet} & \multicolumn{2}{c}{$\geq$3-jet} & \multicolumn{2}{c}{2-jet} & \multicolumn{2}{c}{$\geq$3-jet} \\ \midrule
$Z$~+~jets & 5900 & 100 & 11630 & 160 & 716 & 19 & 2499 & 52 & 84 & 3.2 & 537 & 16\\
$W$~+~jets & 1.29 & 0.13 & 6 & 0.46 & \multicolumn{2}{c}{$<1$} & 1.97 & 0.15 & \multicolumn{2}{c}{$<1$} & \multicolumn{2}{c}{$<1$}\\
Top & 3193 & 57 & 8796 & 87 & 51.9 & 7.1 & 389 & 19 & 1 & 1 & 15.3 & 3.9\\
Diboson & 283 & 47 & 443 & 78 & 83 & 14 & 169 & 30 & 20.4 & 3.5 & 51.8 & 9.7\\
\midrule
Total bkg. & 9378 & 86 & 20880 & 130 & 851 & 19 & 3058 & 44 & 105.5 & 4.1 & 605 & 14\\
\midrule
\midrule
Signal ($\mu=1.02$) & 78 & 14 & 106 & 21 & 34.1 & 6.4 & 59 & 12 & 10.2 & 1.8 & 17.6 & 3.3\\
\midrule
Data
& \multicolumn{1}{c@{\,\hphantom{$\pm$}\,}}{\tablenum[table-alignment=right,table-format=5.0]{9463}}
&& \multicolumn{1}{c@{\,\hphantom{$\pm$}\,}}{\tablenum[table-alignment=right,table-format=5.0]{20927}}
&& \multicolumn{1}{c@{\,\hphantom{$\pm$}\,}}{\tablenum[table-alignment=right,table-format=5.0]{881}}
&& \multicolumn{1}{c@{\,\hphantom{$\pm$}\,}}{\tablenum[table-alignment=right,table-format=5.0]{3148}}
&& \multicolumn{1}{c@{\,\hphantom{$\pm$}\,}}{\tablenum[table-alignment=right,table-format=5.0]{123}}
&& \multicolumn{1}{c@{\,\hphantom{$\pm$}\,}}{\tablenum[table-alignment=right,table-format=5.0]{614}}
\\ \bottomrule
\end{tabular}
\end{sidewaystable}

 
\begin{figure}[htbp]
\begin{center}
\begin{tabular}{cccc}
\includegraphics[width=0.4\linewidth]{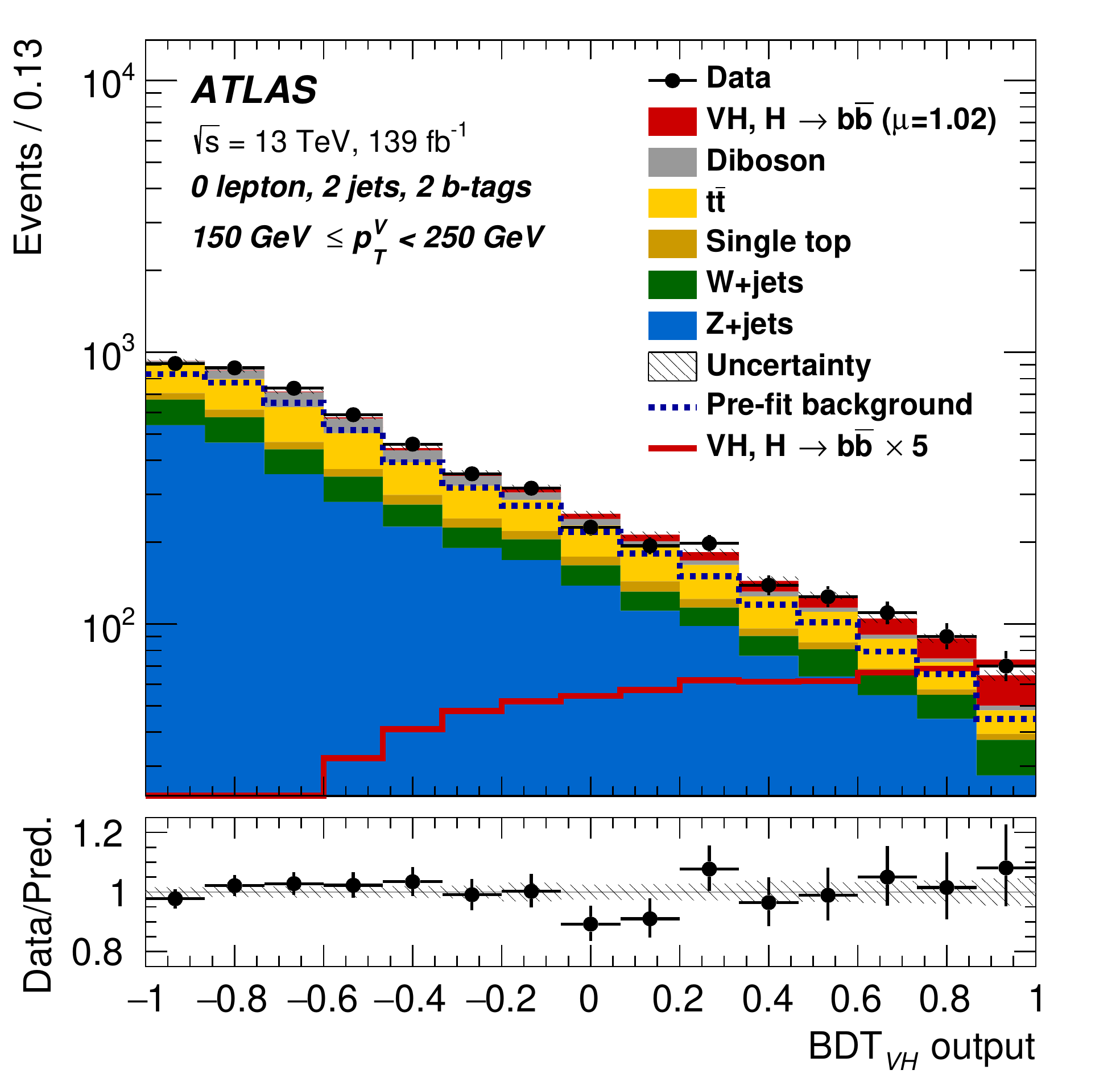}
\includegraphics[width=0.4\linewidth]{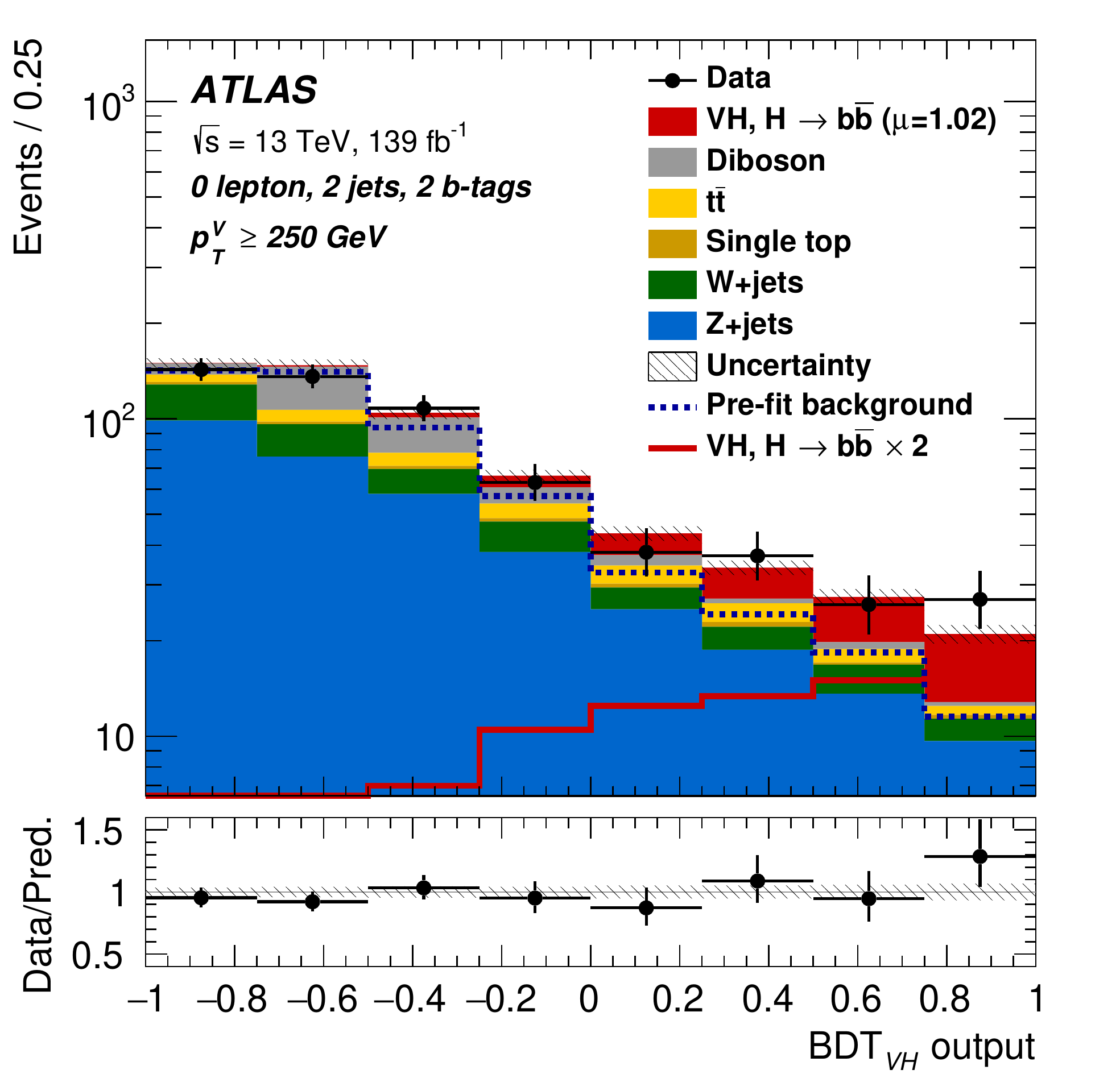}
\\
\includegraphics[width=0.4\linewidth]{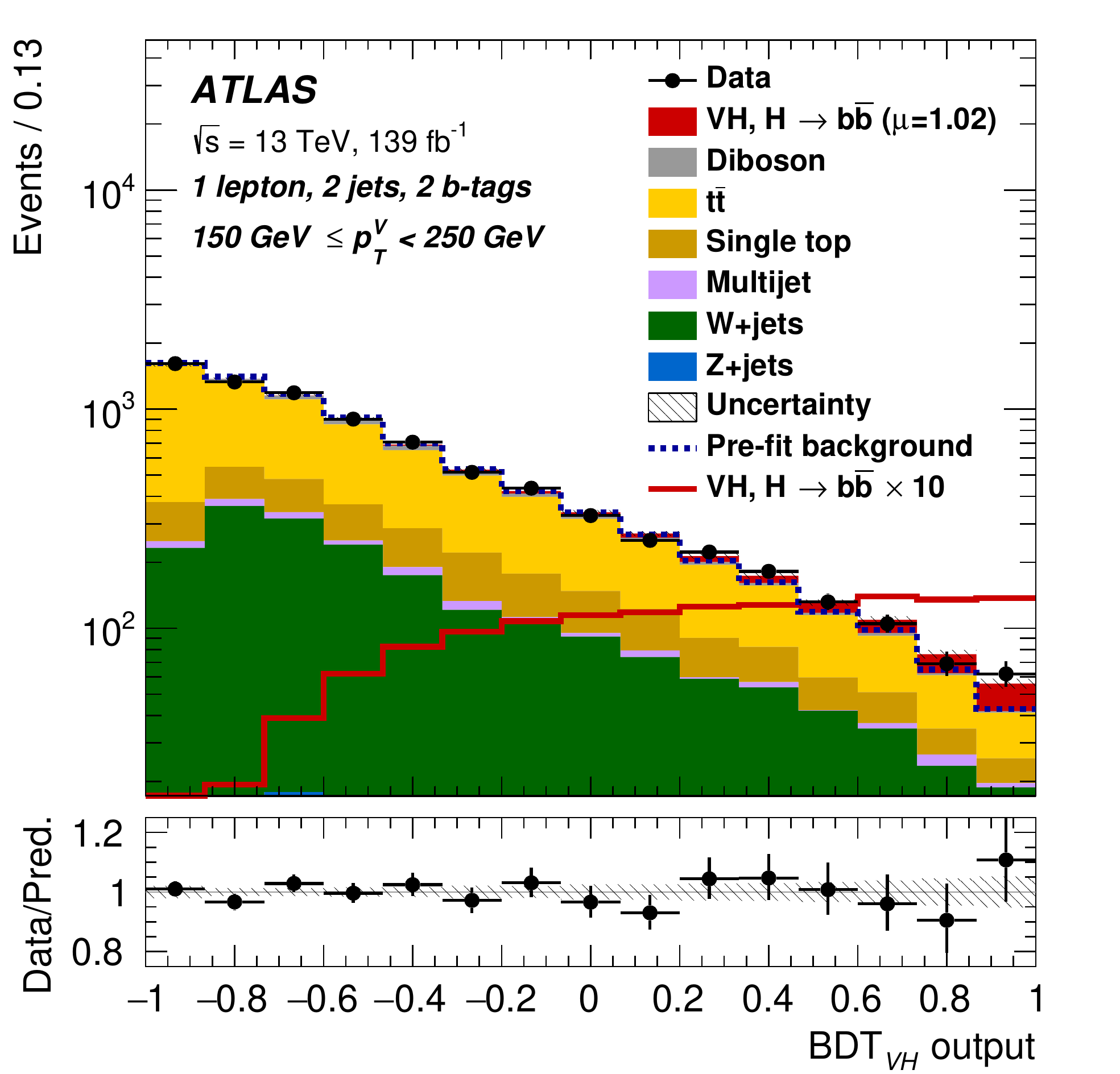}
\includegraphics[width=0.4\linewidth]{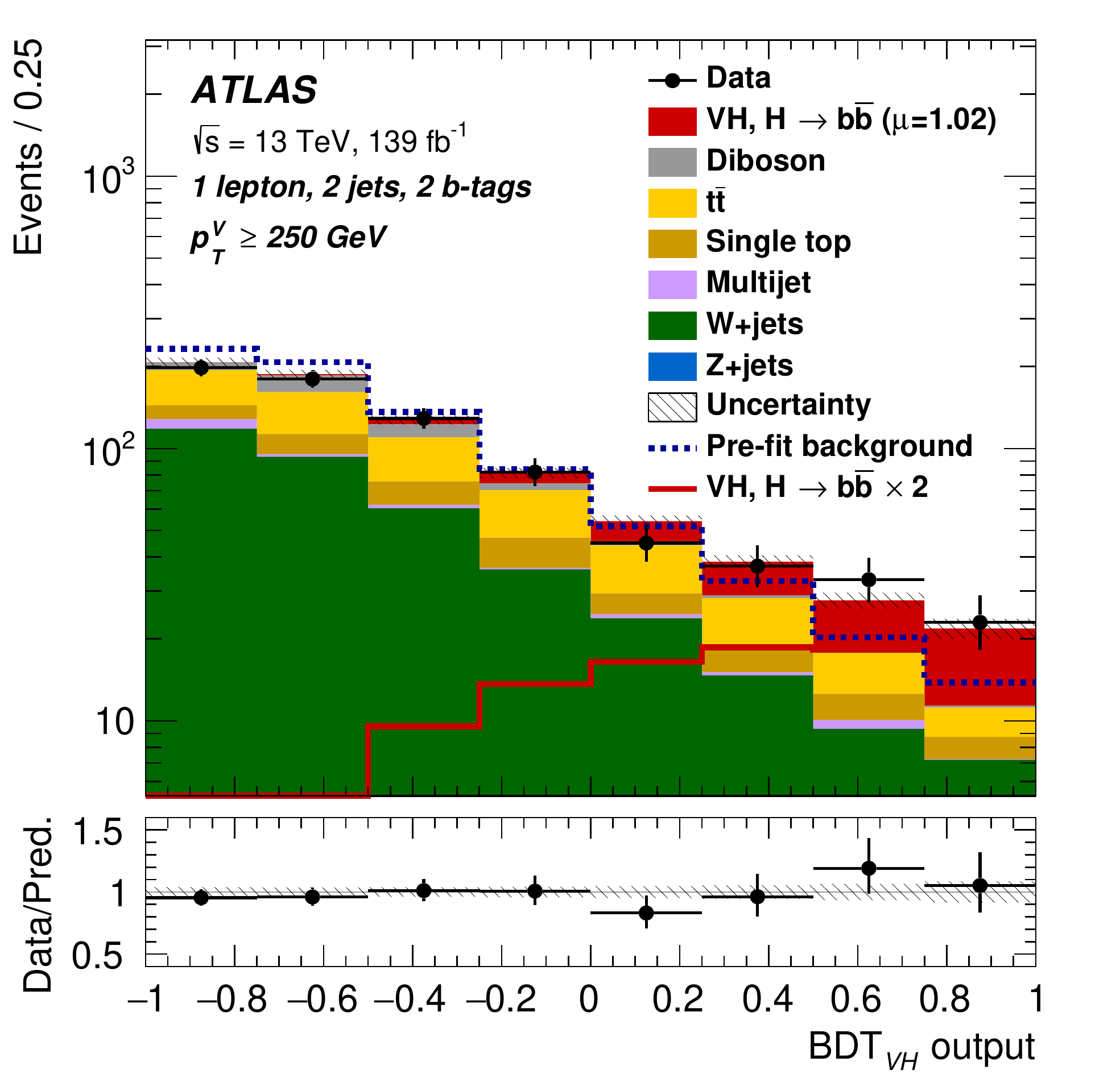}
\\
\includegraphics[width=0.4\linewidth]{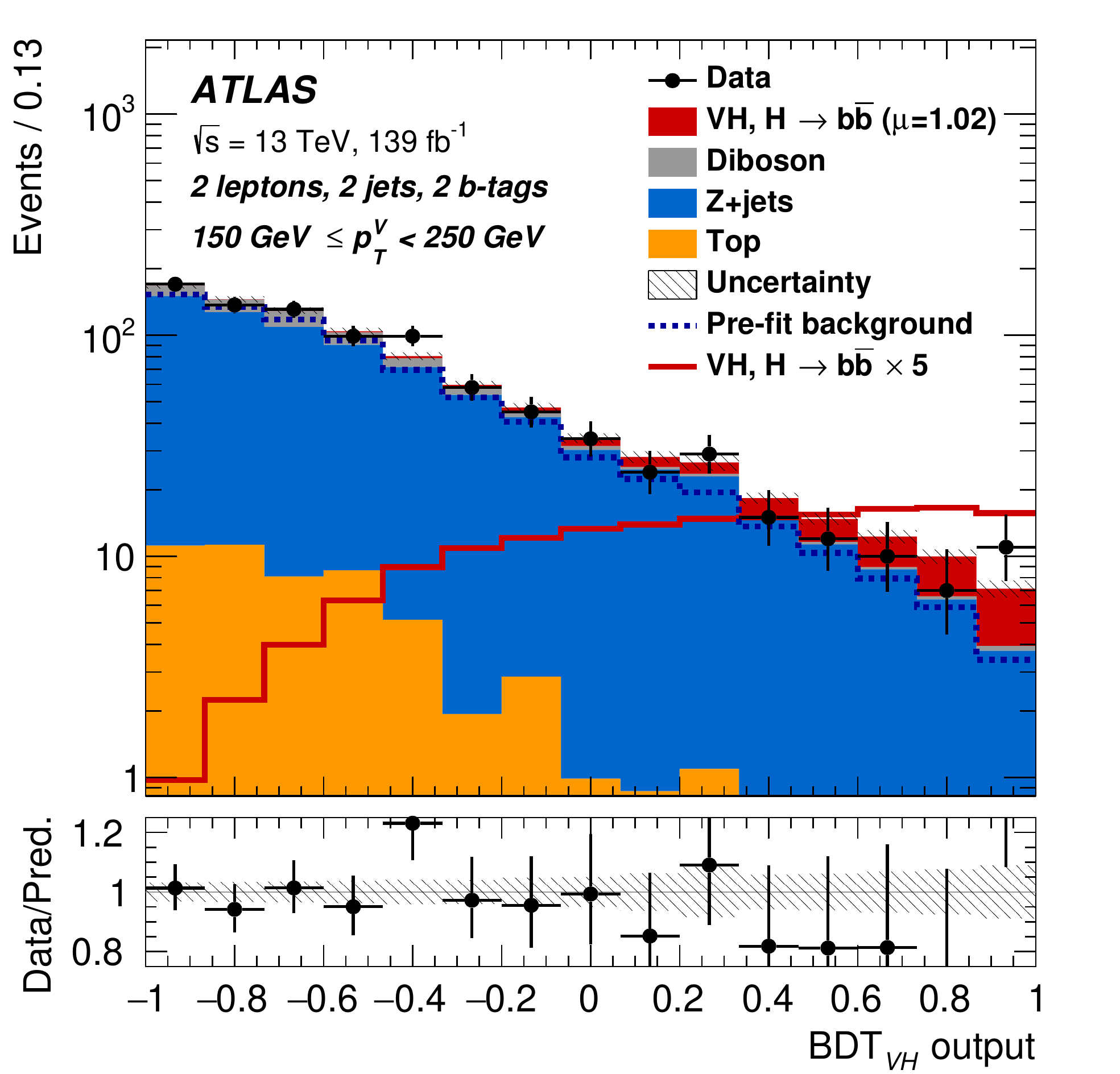}
\includegraphics[width=0.4\linewidth]{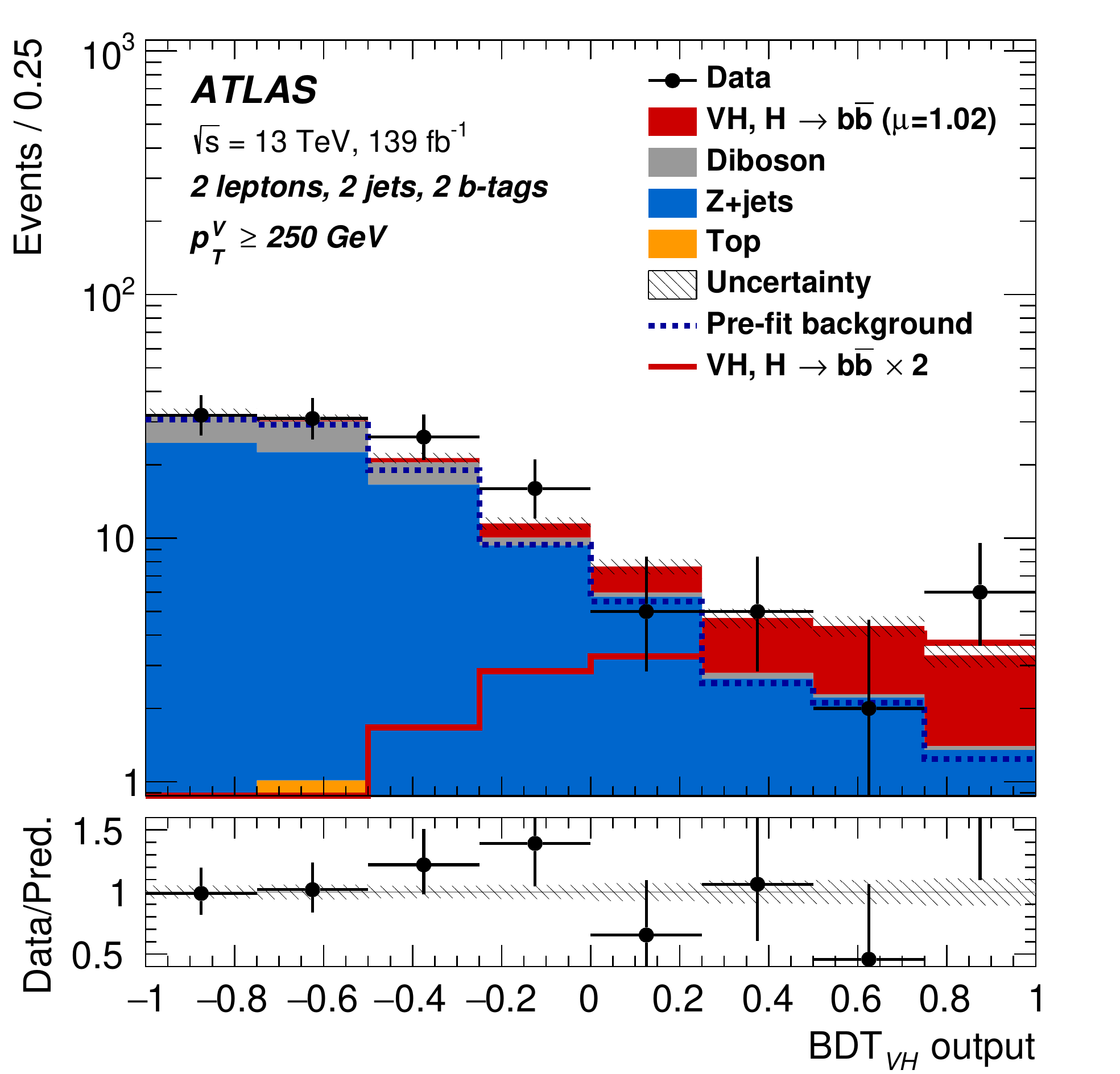}
\\
\end{tabular}
\end{center}
\caption{The BDT$_{VH}$ output post-fit distributions in the 0-lepton~(top), 1-lepton~(middle) and 2-lepton~(bottom) channels for 2-$b$-tag 2-jet events, for the $150<\ptv<250~$\GeV\ (left) and \ptv $>250$~\GeV\ (right) \ptv\ regions. The background contributions after the global likelihood fit are shown as filled histograms. The Higgs boson signal ($\mh = 125$~\GeV) is shown as a filled histogram on top of
the fitted backgrounds normalised to the signal yield extracted from data ($\mu=1.02$),
and unstacked as an unfilled histogram, scaled by the factor indicated in the legend.
The dashed histogram shows the total pre-fit background. The size of the combined statistical and systematic uncertainty for the
sum of the fitted signal and background is indicated by the hatched band. The ratio
of the data to the sum of the fitted signal ($\mu=1.02$) and background is shown in the lower panel.
The BDT$_{VH}$ output distributions are shown with the binning used in the global likelihood fit. \label{fig:mva_output}}
\end{figure}
 
For a Higgs boson mass of 125~\GeV, when all channels are combined, the fitted value of the $VH$ signal strength is:
\begin{eqnarray*}
\mu_{VH}^{bb} = 1.02 ^{+0.18}_{-0.17} = 1.02 ^{+0.12}_{-0.11} \mathrm{(stat.)} ^{+0.14}_{-0.13} \mathrm{(syst.)}.
\end{eqnarray*}
For the $VH$ production mode the background-only hypothesis is rejected with a significance of 6.7 standard deviations, to be compared with an expectation of 6.7 standard deviations~\autocite{Cowan:2010js}.
 
The results of the combined fit when measuring signal strengths separately for
the $WH$ and $ZH$ production processes are shown in
Figure~\ref{fig:mu-higgs-c}. The $WH$ and $ZH$ production modes reject the background-only hypothesis with observed (expected) significances of 4.0 (4.1) and 5.3 (5.1) standard deviations, respectively. The fitted values of the two signal strengths are:
 
\begin{eqnarray*}
\mu_{WH}^{bb} = 0.95 ^{+0.27}_{-0.25} = 0.95 \pm 0.18 \mathrm{(stat.)} ^{+0.19}_{-0.18} \mathrm{(syst.)},\\
\mu_{ZH}^{bb} = 1.08 ^{+0.25}_{-0.23} = 1.08 \pm 0.17 \mathrm{(stat.)} ^{+0.18}_{-0.15} \mathrm{(syst.)},
\end{eqnarray*}
 
with a linear correlation between them of $2.7\%$.
 
\begin{figure}
\begin{center}
\includegraphics[width=0.6\textwidth]{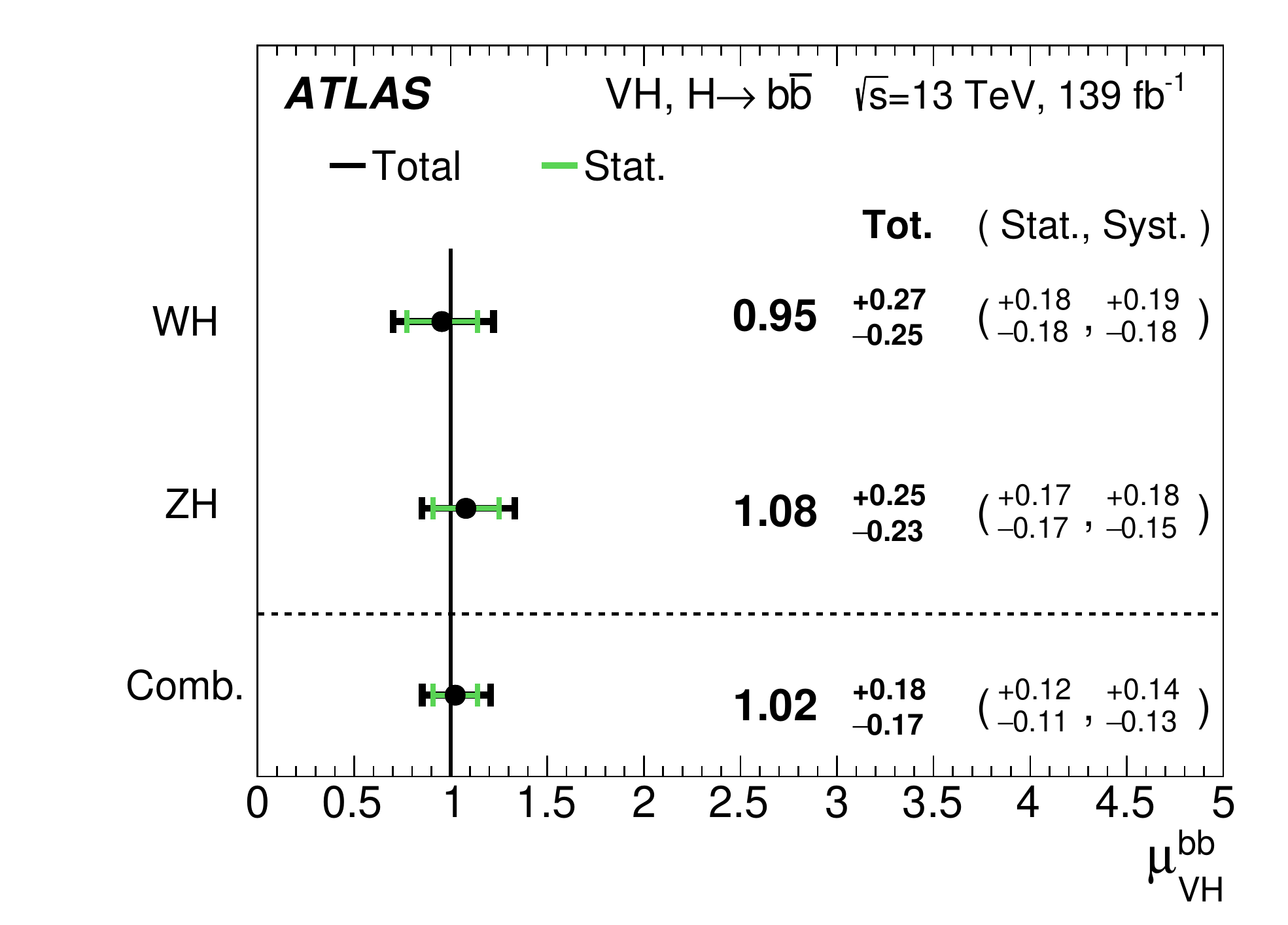}
\end{center}
\caption{The fitted values of the Higgs boson signal strength $\mu_{VH}^{bb}$ for $\mh=125$~\GeV\ for the $WH$ and $ZH$ processes and their combination. The individual $\mu_{VH}^{bb}$ values for the $(W/Z)H$ processes
are obtained from a simultaneous fit with the signal strength for each of the $WH$ and $ZH$ processes floating independently.
The probability of compatibility of the individual signal strengths is 71\%.\label{fig:mu-higgs-c}}
\end{figure}
 
The effects of systematic uncertainties on the measurement of the $VH$, $WH$ and $ZH$ signal strengths are
displayed in Table~\ref{tab:syst_unc}.
The impact of a set of systematic uncertainties is defined as the
difference in quadrature between the uncertainty in $\mu$ computed when
all NPs are fitted and that when the NPs in the set are fixed to
their best-fit values.
The total statistical
uncertainty is defined as the uncertainty in $\mu$ when all the NPs are fixed to their
best-fit values. The total systematic uncertainty is then defined as the difference in
quadrature between the total uncertainty in $\mu$ and the total statistical uncertainty.
For the $WH$ and $ZH$ signal strength measurements the total statistical and systematic uncertainties are similar in size, with the $b$-tagging, jet, \met, background modelling and signal systematic uncertainties all making important contributions to the total systematic uncertainty.
The impact of the statistical uncertainty from the simulated event samples has been significantly reduced compared to the previous result~\cite{HIGG-2018-50}, due to the measures taken to considerably enhance the number of simulated events.
 
\begin{table}[tb]
\caption{Breakdown of the contributions to the uncertainty in $\mu_{VH}^{bb}$ for the $VH$, $WH$ and $ZH$ signal strength measurements. The sum in quadrature of the systematic uncertainties attached to the
categories differs from the total systematic uncertainty due to correlations.
\label{tab:syst_unc}}
\begin{center}
\begin{tabular}{ l | l  c | c c}
\toprule
\multicolumn{2}{l}{\multirow{2}{*}{Source of uncertainty}} & \multicolumn{3}{c}{$\sigma_\mu$}  \\
\multicolumn{2}{l}{} & $VH$ & $WH$ & $ZH$ \\
\midrule
\multicolumn{2}{l}{Total} &  $  0.177 $ &  $  0.260 $ &  $  0.240 $ \\
\multicolumn{2}{l}{Statistical} &  $  0.115 $ &  $  0.182 $ &  $  0.171 $ \\
\multicolumn{2}{l}{Systematic} & $  0.134 $ &  $  0.186 $ &  $  0.168 $ \\
\midrule
\multicolumn{4}{l}{Statistical uncertainties}\\
\midrule
\multicolumn{2}{l}{Data statistical} & $  0.108 $ &  $  0.171 $ &  $  0.157 $ \\
\multicolumn{2}{l}{\ttbar $e\mu$ control region} & $  0.014 $ &  $  0.003 $ &  $  0.026 $ \\
\multicolumn{2}{l}{Floating normalisations} & $  0.034 $ &  $  0.061 $ &  $  0.045 $ \\
\midrule
\multicolumn{4}{l}{Experimental uncertainties}\\
\midrule
\multicolumn{2}{l}{Jets} & $  0.043 $ &  $  0.050 $ &  $  0.057 $ \\
\multicolumn{2}{l}{\met} & $  0.015 $ &  $  0.045 $ &  $  0.013 $ \\
\multicolumn{2}{l}{Leptons} & $  0.004 $ &  $  0.015 $ &  $  0.005 $ \\
\multirow{3}{*}{$b$-tagging~~~}
& $b$-jets &  $  0.045 $ &  $  0.025 $ &  $  0.064 $ \\
& $c$-jets &  $  0.035 $ &  $  0.068 $ &  $  0.010 $ \\
& light-flavour jets & $  0.009 $ &  $  0.004 $ &  $  0.014 $ \\
\multicolumn{2}{l}{Pile-up} & $  0.003 $ &  $  0.002 $ &  $  0.007 $ \\
\multicolumn{2}{l}{Luminosity} & $  0.016 $ &  $  0.016 $ &  $  0.016 $ \\
\midrule
\multicolumn{4}{l}{Theoretical and modelling uncertainties}\\
\midrule
\multicolumn{2}{l}{Signal} & $  0.072 $ &  $  0.060 $ &  $  0.107 $ \\
\multicolumn{4}{l}{} \\
\multicolumn{2}{l}{$Z$~+~jets} & $  0.032 $ &  $  0.013 $ &  $  0.059 $ \\
\multicolumn{2}{l}{$W$~+~jets} & $  0.040 $ &  $  0.079 $ &  $  0.009 $ \\
\multicolumn{2}{l}{\ttb} & $  0.021 $ &  $  0.046 $ &  $  0.029 $ \\
\multicolumn{2}{l}{Single top quark} & $  0.019 $ &  $  0.048 $ &  $  0.015 $ \\
\multicolumn{2}{l}{Diboson} & $  0.033 $ &  $  0.033 $ &  $  0.039 $ \\
\multicolumn{2}{l}{Multi-jet} & $  0.005 $ &  $  0.017 $ &  $  0.005 $ \\
\multicolumn{4}{l}{} \\
\multicolumn{2}{l}{MC statistical} & $  0.031 $ &  $  0.055 $ &  $  0.038 $ \\
\bottomrule
\end{tabular}
\end{center}
\end{table}
 
\FloatBarrier
 
\subsubsection{Dijet-mass cross-check}
 
From the fit to \mbb, for all channels combined, the value of the signal strength is
\begin{eqnarray*}
\mu_{VH}^{bb} = 1.17^{+0.25}_{-0.23} = 1.17 \pm 0.16 \mathrm{(stat.)} ^{+0.19}_{-0.16} \mathrm{(syst.)}.
\end{eqnarray*}
Using the `bootstrap' method~\cite{HIGG-2013-23}, the dijet-mass and nominal multivariate analysis results are found to be statistically compatible at the level of 1.1 standard deviations.
The observed excess rejects the background-only hypothesis with a significance of 5.5 standard deviations,
compared to an expectation of 4.9 standard deviations.
Good agreement is also found when comparing the values of signal strengths in the individual channels
from the dijet-mass analysis with those from the multivariate analysis.
 
The \mbb\ distribution is shown in Figure~\ref{fig:mbb_cutbased} summed over all channels and
regions, weighted by their respective values of the ratio of fitted Higgs boson
signal to background yields and after subtraction of all backgrounds except for the
$WZ$ and $ZZ$ diboson processes.
 
\begin{figure}[hbp!]
\begin{center}
\includegraphics[width=0.59\textwidth]{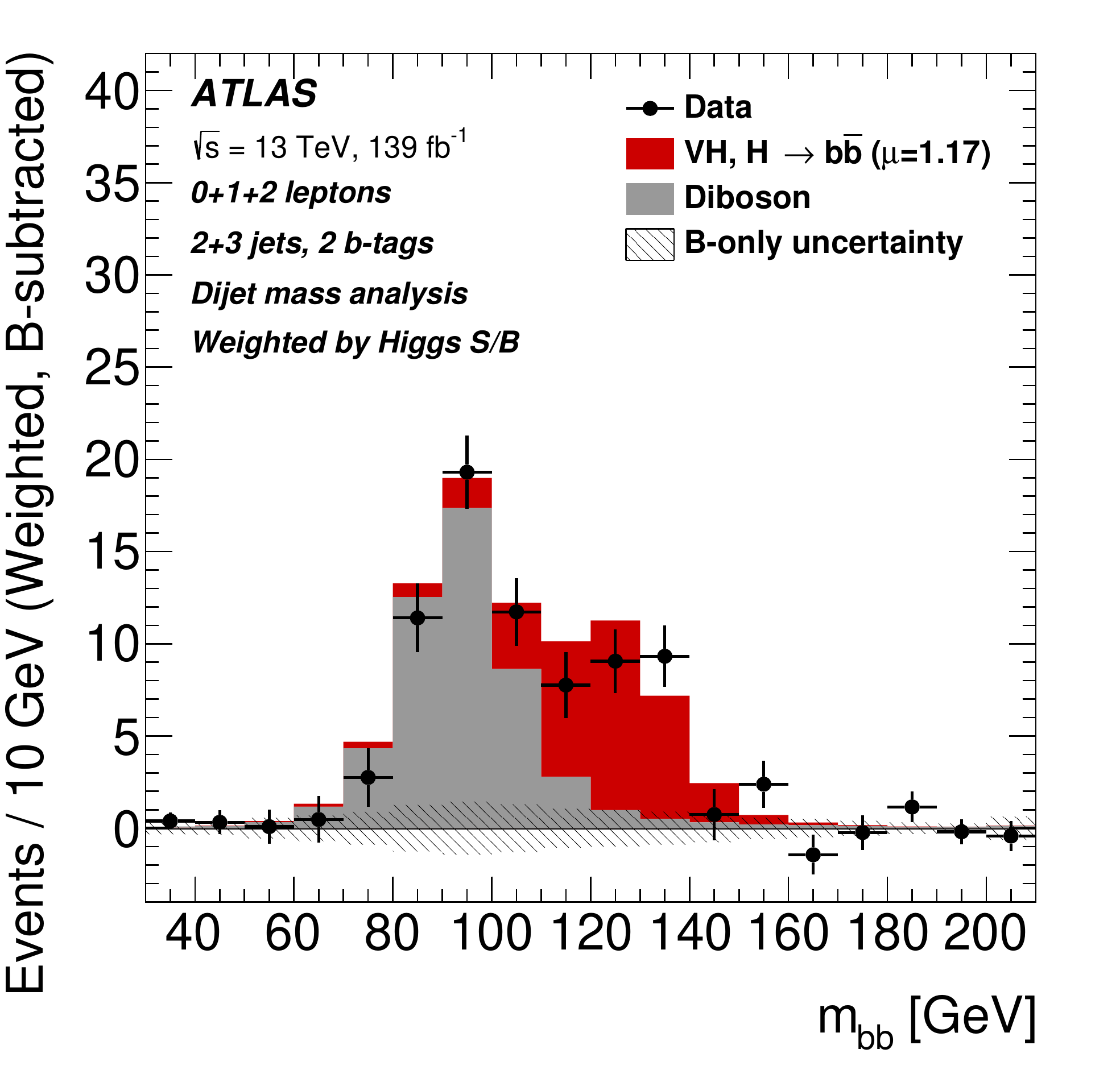}
\end{center}
\caption{The distribution of \mbb\ in data after subtraction of all backgrounds
except for the $WZ$ and $ZZ$ diboson processes,
as obtained with the dijet-mass analysis.
The contributions from all lepton channels, \ptv\ regions and
number-of-jets categories are summed and weighted by their respective $S/B$ ratios,
with $S$ being the total fitted signal and $B$ the total fitted background in each region.
The expected contribution of the associated $WH$ and $ZH$ production of a SM Higgs boson with $\mh=125$~\GeV\
is shown scaled by the measured signal strength ($\mu = 1.17$).
The size of the combined statistical and systematic uncertainty
for the fitted background is indicated by the hatched band.
\label{fig:mbb_cutbased}}
\end{figure}
 
\subsubsection{Diboson validation}
 
The measurement of $VZ$ production using a multivariate approach, as a validation of the Higgs boson analysis, returns a signal strength of
\begin{eqnarray*}
\mu^{bb}_{VZ} = 0.93 ^{+0.16}_{-0.13} = 0.93 ^{+0.07}_{-0.06} \mathrm{(stat.)} ^{+ 0.14}_{-0.12} \mathrm{(syst.)},
\end{eqnarray*}
 
in good agreement with the Standard Model prediction. Analogously to the nominal analysis, fits are also performed with separate signal strengths for the $WZ$ and $ZZ$ production modes, and the results are shown in Figure~\ref{fig:mu-higgs-c-dib}.
 
\begin{figure}[tbhp!]
\begin{center}
\includegraphics[width=0.7\textwidth]{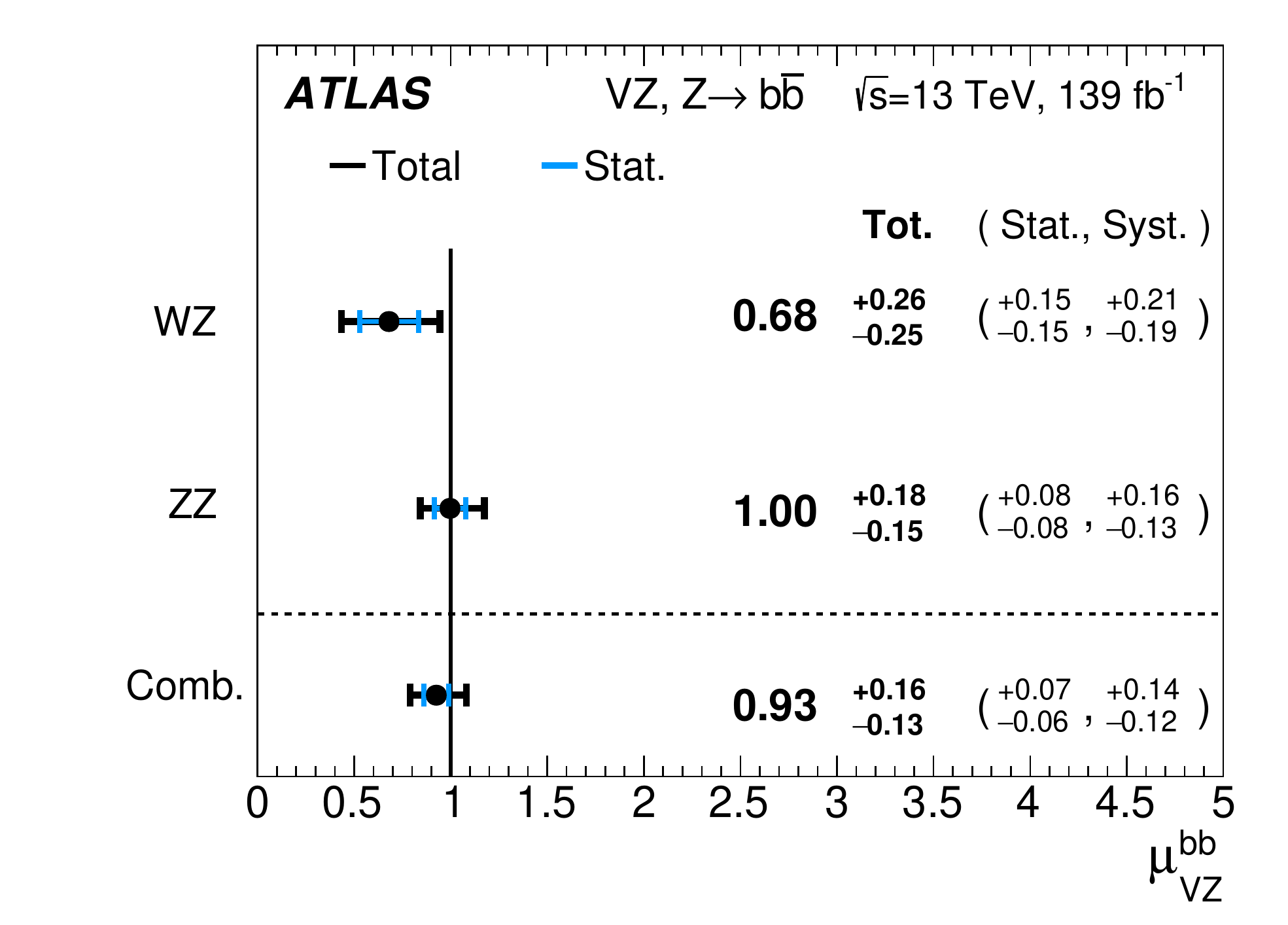}
\end{center}
\caption{The fitted values of the $VZ$  signal strength $\mu_{VZ}^{bb}$ for the $WZ$ and $ZZ$ processes
and their combination. The individual $\mu_{VZ}^{bb}$ values for the $WZ$ and $ZZ$ processes
are obtained from a simultaneous fit with the signal strengths for each of the $WZ$ and $ZZ$ processes floating independently.
The probability of compatibility of the individual signal strengths is 27\%.
\label{fig:mu-higgs-c-dib}}
\end{figure}
 
\subsection{Cross-section measurements}
 
The measured $VH$ cross-sections times
the $H\to b\bar{b}$ and $V\to$~leptons branching fractions,
$\sigma\times B$, together with the SM predictions in the reduced STXS regions, are summarised in
Table~\ref{tab:XS3and5POI} and Figure~\ref{fig:Xsections}.
The cross-sections are all consistent with the Standard Model expectations and are measured with relative uncertainties varying from 30\% in the highest \ptv\ region to 85\% in the lowest \ptv\ region. The data statistical uncertainty is the largest single uncertainty in all regions, although in the lower \ptv\ regions systematic uncertainties make a sizeable contribution to the total uncertainty. In all regions there are large contributions from the background modelling, $b$-tagging and jet systematic uncertainties. In the lowest \ptv\ region in both the $WH$ and $ZH$ measurements, the \met\ uncertainty is one of the largest uncertainties. For the $ZH$ measurements, the signal uncertainties also make a sizeable contribution due to the limited precision of the theoretical calculations of the $gg\to ZH$ process.
 
\begin{table}[!htbp]
\setlength{\extrarowheight}{4pt}
\begin{center}
\sisetup{round-mode=places, round-precision=0, group-digits=integer, group-minimum-digits=6}
\caption{Best-fit values and uncertainties for the $VH$,
$V\to$~leptons cross-section
times the $H\to b\bar{b}$ branching fraction, in the reduced STXS scheme.
The SM predictions for each region, computed using the inclusive
cross-section
calculations and the simulated event samples are also shown.
The contributions to the total uncertainty in the measurements from
statistical (Stat.~unc.) or systematic uncertainties (Syst.~unc.) in
the signal prediction (Th.~sig.), background prediction (Th.~bkg.),
and in experimental performance (Exp.) are given separately.
The total systematic uncertainty, equal to the difference in quadrature between
the total uncertainty and the statistical uncertainty, differs from the
sum in quadrature of the Th.~sig., Th.~bkg., and Exp.\ systematic
uncertainties due to correlations.
All leptonic decays of the $V$ bosons (including those to $\tau$-leptons, $\ell=e,\mu,\tau$) are considered. These results along with the corresponding correlation matrix are available in the HEPData repository~\cite{hepdata}.}
\label{tab:XS3and5POI}
\resizebox{\textwidth}{!}{
\begin{tabular}{cc  rcl  rcr | cr  cr  cr  cr }
\toprule
\multicolumn{2}{c}{STXS region}                                      & \multicolumn{3}{c}{SM prediction}  & \multicolumn{3}{c|}{Result} & \multicolumn{2}{c}{Stat. unc.}    & \multicolumn{6}{c}{Syst. unc. [fb]}                     \\
\cline{10-15}
Process & $p_{\textrm{T}}^{V\textrm{, t}}$ interval & \multicolumn{3}{c}{[fb]}           & \multicolumn{3}{c|}{[fb]}   & \multicolumn{2}{c}{[fb]}           & \multicolumn{2}{c}{Th. sig.}           & \multicolumn{2}{c}{Th. bkg.}            & \multicolumn{2}{c}{Exp.}           \\
\midrule
$W(\ell\nu)H$ &  150--250~\GeV        & $24.0$ & $\pm$ & $1.1$        & $19.0$ & $\pm$ & $12.1$ &  $\pm$ & $7.7$  & $\pm$ & 0.9 & $\pm$ &  5.5  & $\pm$ & 6.0  \\
$W(\ell\nu)H$ &  $>250$~\GeV             & $7.1$  & $\pm$ & $0.3$        & $7.2$  & $\pm$ & $2.2$  &  $\pm$ & $1.9$  & $\pm$ & 0.4 & $\pm$ &  0.8  & $\pm$ & 0.7  \\
$Z(\ell\ell/\nu\nu)H$ &  75--150~\GeV    & $50.6$ & $\pm$ & $4.1$        & $42.5$ & $\pm$ & $35.9$  & $\pm$ & $25.3$ & $\pm$ & 5.6 & $\pm$ & 17.2 & $\pm$ & 19.7 \\
$Z(\ell\ell/\nu\nu)H$ &  150--250~\GeV    & $18.8$ & $\pm$ & $2.4$        & $20.5$ & $\pm$ & $6.2$  &  $\pm$ & $5.0$  & $\pm$ & 2.3  & $\pm$ &  2.4  & $\pm$ & 2.3  \\
$Z(\ell\ell/\nu\nu)H$ &  >250~\GeV        & $4.9$  & $\pm$ & $0.5$        & $5.4$ & $\pm$ & $1.7$  &  $\pm$ & $1.5$  & $\pm$ & 0.5  & $\pm$ &  0.5  & $\pm$ & 0.3  \\
\bottomrule
\end{tabular}
}
\end{center}
\end{table}
 
\begin{figure}[!htbp]
\begin{center}
\includegraphics[width=0.7\textwidth]{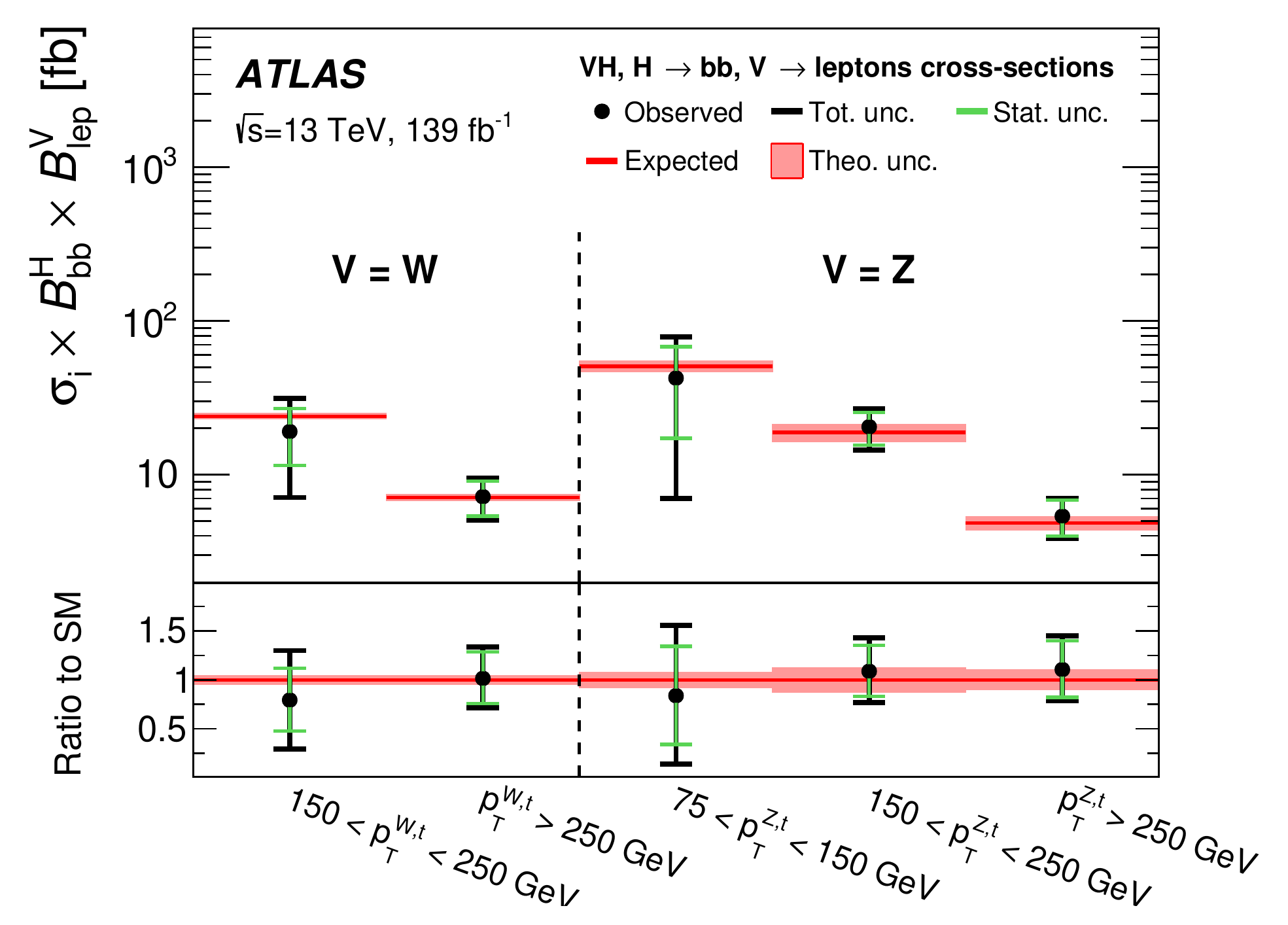}
\caption{Measured $VH$, $V\to$~leptons cross-sections times the $H\to b\bar{b}$ branching fraction in the reduced STXS scheme. }
\label{fig:Xsections}
\end{center}
\end{figure}
 
 
\section{Constraints on effective interactions}
The strength and tensor structure of the process $VH, H \to b\bar{b}$
are investigated using an effective Lagrangian
approach.
Extra terms are added to the SM Lagrangian ($\mathcal{L}_{\mathrm{SM}}$) to obtain
an effective Lagrangian ($\mathcal{L}_{\mathrm{SMEFT}}$) following the approach in Refs.~\cite{Contino:2013kra, Brivio:2017btx}:
 
\begin{equation}\nonumber
\mathcal{L}_{\mathrm{SMEFT}} = \mathcal{L}_{\mathrm{SM}} + \sum_i \frac{c_i^{(D)}}{\Lambda^{D-4}} \mathcal{Q}_i^{(D)},
\end{equation}
 
where $\Lambda$ is the energy scale of the new interactions,
${Q}_i^{(D)}$ are dimension-$D$ operators, and
$c_i^{(D)}$ are numerical Wilson coefficients.
Only $D=6$ operators are considered in this study,
since $D=5$ and $D=7$ operators violate lepton or baryon number,
whilst $D>7$ operators are further suppressed by powers of $\Lambda$.
 
The STXS measurements are used to constrain the coefficients of the operators in the `Warsaw' formulation~\cite{Grzadkowski:2010es}, which provides a complete set of independent operators when considering those allowed by the SM gauge symmetries. Thirteen operators directly affect the $VH$ cross-section~\cite{ATL-PHYS-PUB-2019-042}. This analysis has significant sensitivity to the six operators detailed in Table~\ref{tab:eftoperators}, in addition to the operator which directly affects the $\Hbb$ decay width.

\begin{table}[tph]
\caption{Wilson coefficients $c_i$ and corresponding dimension-6 SMEFT operators $\mathcal{Q}_i$, to which this analysis is sensitive, in the Warsaw formulation~\cite{Grzadkowski:2010es}.}
\label{tab:eftoperators}
\begin{center}
\begin{tabular}{cccc}
\toprule
Wilson coefficient & Operator & \multicolumn{2}{c}{Impacted vertex}  \\[4pt]
& & Production & Decay \\[4pt]
\midrule
$c_{HWB}$       & $\mathcal{Q}_{HWB} =$ $H^\dagger \tau^I H W_{\mu\nu}^I B^{\mu\nu}$ & $HZZ$ &  \\[4pt]
$c_{HW}$        & $\mathcal{Q}_{HW} =$ $H^\dagger H W^I_{\mu\nu} W^{\mu\nu}_I$ & $HZZ, HWW$ &  \\[4pt]
$c_{Hq}^{(3)}$  & $\mathcal{Q}_{Hq}^{(3)} =$ $(H^\dagger i \overleftrightarrow{D_{\mu}^I} H)(\bar{q}_p \tau^I \gamma^\mu q_r)$ & $qqZH, qq'WH$ &  \\[4pt]
$c_{Hq}^{(1)}$  & $\mathcal{Q}_{Hq}^{(1)} =$ $(H^\dagger i \overleftrightarrow{D_{\mu}} H)(\bar{q}_p \gamma^\mu q_r)$ & $qqZH$ &  \\[4pt]
$c_{Hu}$        & $\mathcal{Q}_{Hu} =$ $(H^\dagger i \overleftrightarrow{D_{\mu}} H)(\bar{u}_p \gamma^\mu u_r)$ & $qqZH$ &  \\[4pt]
$c_{Hd}$        & $\mathcal{Q}_{Hd} =$ $(H^\dagger i \overleftrightarrow{D_{\mu}} H)(\bar{d}_p \gamma^\mu d_r)$ & $qqZH$ &  \\[4pt]
$c_{dH}$        & $\mathcal{Q}_{dH} =$ $(H^\dagger H)(\bar{q} d H)$ &  &$Hbb$  \\
\bottomrule
\end{tabular}
\end{center}
 
\end{table}

Following methodologies similar to those outlined in Ref.~\cite{Brivio:2017btx}, a parameterisation of the STXS production cross-section and Higgs boson decay rates in terms of the SMEFT parameters is derived, in this case based upon leading-order predictions made using the SMEFTsim package~\cite{Brivio:2017btx}. 
The interference terms between the
SM and BSM amplitudes are linear in the coefficients
and of order $1/\Lambda^2$, while BSM contributions
are quadratic in the coefficients and of order $1/\Lambda^4$.
Linear terms from $D=8$ operators are suppressed by the same $1/\Lambda^4$ factor as the quadratic $D=6$ terms. However, it is currently not possible to include such terms, so results for both the linear and linear plus quadratic $D=6$ terms are studied to provide some indication of the effect $D=8$ linear terms could have on the result.
Modifications of the $gg\to ZH$ production cross-section are only introduced
by either higher-dimension ($D \geq 8$) operators or corrections
that are formally at NNLO in QCD, and are
not included in this study. The expected $gg\to ZH$
contribution is fixed to the SM prediction within uncertainties. The dependence of the experimental acceptance in each analysis region on the Wilson coefficients is not accounted for in this study, although it was verified that the impact on the acceptance from the EFT operators was at most 10\%.
 
Maximum-likelihood fits across the STXS
regions are performed to determine the Wilson coefficients.
All coefficients but one are assumed to vanish, and one-dimensional confidence level (CL) intervals are
inferred for the coefficient under study both with and without the quadratic terms.
An example negative-log-likelihood one-dimensional projection
is shown in Figure~\ref{fig:EFT_1Dscan} for $c_{Hq}^{(3)}$, and the
68\% and 95\% CL intervals are summarised in Figure~\ref{fig:1d_limits_summary_main} for the four coefficients to which the analysis has greatest sensitivity, in addition to the $c_{dH}$ coefficient which directly affects the $H\to b\bar{b}$ decay width. As detailed in Table~\ref{tab:eftoperators}, the $\mathcal{Q}_{Hu}$, $\mathcal{Q}_{Hd}$ and $\mathcal{Q}_{Hq}^{(1)}$ operators have a similar impact and as such are found to be highly degenerate, so only a representative result for $\mathcal{Q}_{Hu}$ is shown. The coefficient $c_{Hq}^{(3)}$ is constrained
at 68\%~CL to be no more than a few percent, whilst the constraints on the other three coefficients range from 10\%--30\% to order unity and $c_{dH}$ has much weaker constraints. In most cases the observed constraints are found to significantly depend on the presence of the quadratic terms, indicating that $D=8$ linear terms could also have a non-negligible effect.
 
These limits were also produced using the full likelihood and using only the STXS measurement central values and covariance matrix. It was found that the two methods produced results that are consistent with each other within $\sim10\%$--$20$\% for the majority of operators and to within $\sim30\%$ for the two operators with the weakest constraints, $\mathcal{Q}_{dH}$ and $\mathcal{Q}_{HWB}$.
 
\begin{figure}[!ht]
\begin{center}
\includegraphics[width=0.5\linewidth]{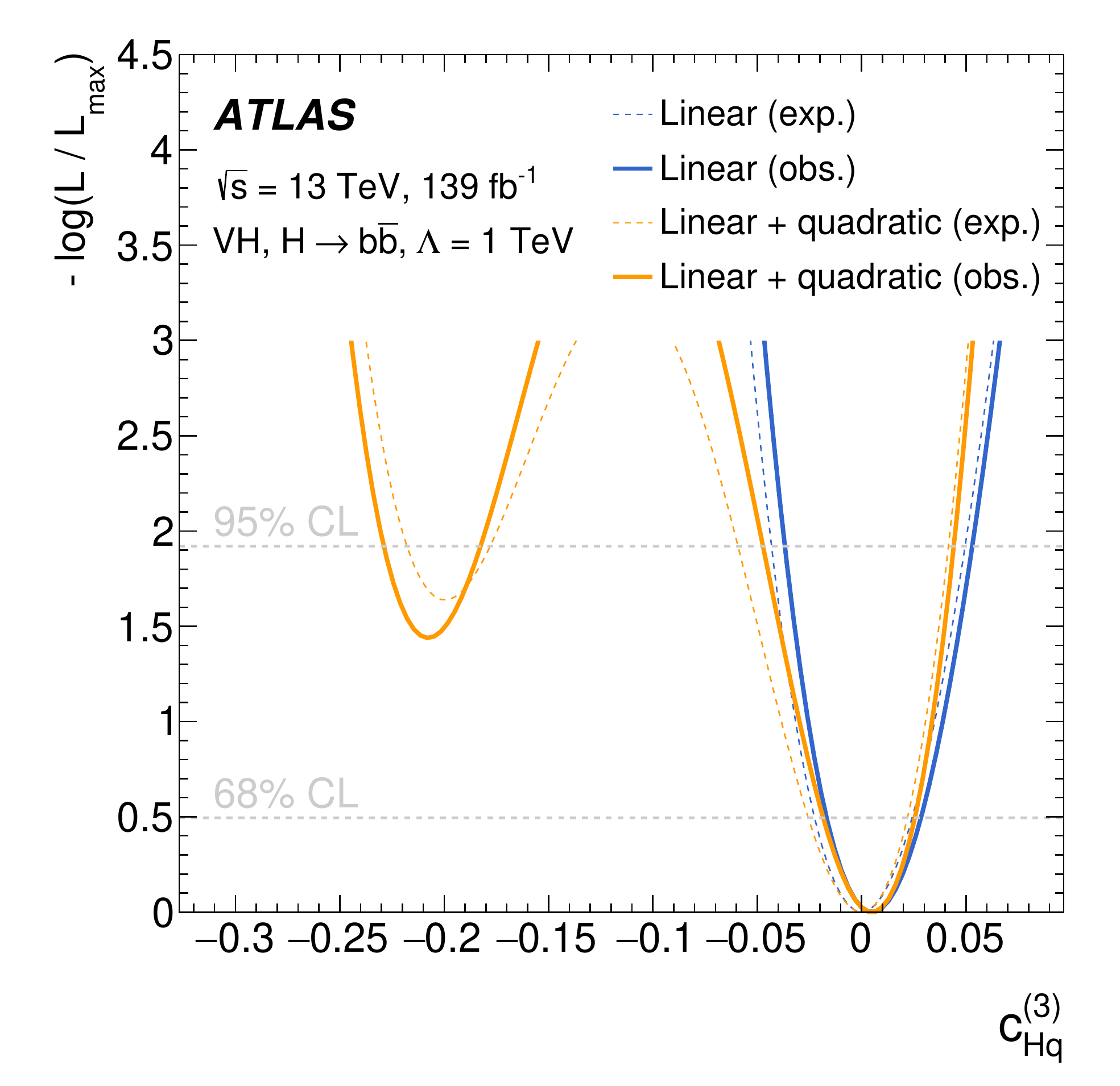}
\end{center}
\caption{The observed (solid) and expected (dotted) profiled
negative-log-likelihood functions for the one-dimensional
fits to constrain the coefficient $c_{Hq}^{(3)}$
of an effective Lagrangian when the other coefficients are assumed to vanish, shown for the case where only linear (blue) or linear and quadratic (orange) terms are considered.}
\label{fig:EFT_1Dscan}
\end{figure}
 
\begin{figure} [!h]
\begin{center}
\includegraphics[width=0.5\linewidth]{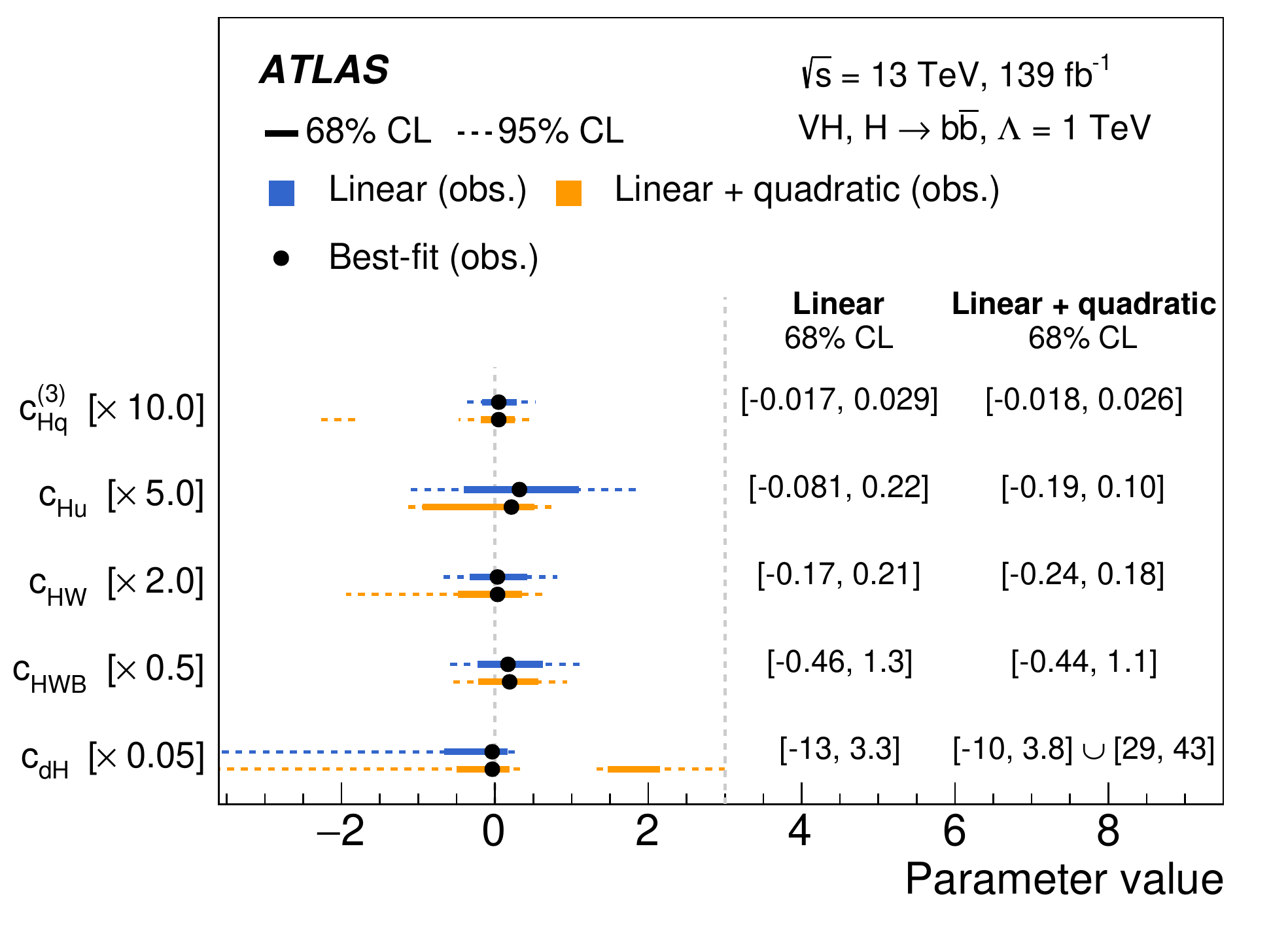}
\end{center}
\caption{Summary of the observed best-fit values and one-dimensional confidence intervals for the Wilson coefficients of the Warsaw-basis operators to which this analysis has the greatest sensitivity along with the $c_{dH}$ coefficient which directly affects the $H\to b\bar{b}$ decay width. Limits are shown for the case where only linear (blue) or linear and quadratic (orange) terms are considered and confidence intervals are shown at both 68\% CL (solid lines) and 95\% CL (dashed lines).}
\label{fig:1d_limits_summary_main}
\end{figure}
 
As there are only five STXS regions, attempting to simultaneously extract constraints on multiple coefficients, some of which have similar effects, leads to unmanageable correlations. An alternative approach is to fit an orthogonal set of linear combinations of the Wilson coefficients of the Warsaw-basis operators. This removes the assumption, inherent in the one-dimensional limits, that only one operator acts at a time. Based upon the procedure outlined in Ref.~\cite{ATL-PHYS-PUB-2019-042}, eigenvectors are determined from the Hessian matrix of the STXS likelihood fit to data, after it has been re-expressed in terms of the Wilson coefficients. This approach only considers the linear terms and the $H \to b\bar{b}$ partial width, with a dedicated independent parameter added to account for the modifications to the total width.
 
The resulting five eigenvectors are shown in Table~\ref{tabular:eigenvectors_data}. They are labelled as $\mathrm{E0}$--$\mathrm{E4}$  and ordered in terms of experimental sensitivity, with $\mathrm{E0}$ having the greatest and $\mathrm{E4}$ the least. The eigenvectors contain information about the sensitivity of the analysis to degenerate deformations of the SM. The leading eigenvector, $\mathrm{E0}$, consists almost exclusively of $c_{Hq}^{(3)}$, which is also the coefficient most constrained in the one-dimensional limits, with similar limits obtained in both cases. The second eigenvector, $\mathrm{E1}$, is dominated by $c_{Hu}$, but has sizeable contributions from $c_{Hd}$ and $c_{Hq}^{(1)}$, suggesting only a linear combination of these coefficients can be constrained given the degeneracy between them. The eigenvector $\mathrm{E2}$ demonstrates sensitivity to a combination of the branching ratio and $c_{HW}$, whilst $\mathrm{E3}$  has limited sensitivity to a combination of $c_{HWB}$ and $c_{Hq}^{(1)}$. The analysis has negligible sensitivity to the fifth eigenvector. Figure~\ref{fig:eigenoperator_fit} shows the impact on the STXS cross-section measurements when varying the coefficients for the four leading eigenvectors within their $1\sigma$ bounds. The analysis has greatest sensitivity to coefficients which predominantly increase the cross-section in the higher $\ptv$ STXS regions (E0 and E1), with lower sensitivity to those which predominantly impact the lower $\ptv$ STXS regions (E2 and E3).
 
\begin{table}[htbp]
\caption{The composition and eigenvalues of the eigenvectors, which are composed of a linear combination of the Wilson coefficients of the Warsaw-basis operators~\cite{Grzadkowski:2010es}. All modifications that alter the branching ratio are absorbed into an additional independent term ($\Delta\mathrm{BR}/\mathrm{BR}_{\mathrm{SM}}$), which linearly alters the branching ratio and all contributions with a coefficient below 0.2 are omitted. The full composition of the eigenvectors is available in the HEPData repository~\cite{hepdata}.}
\label{tabular:eigenvectors_data}
\begin{center}
\begin{tabular}{c c c}
\toprule
Wilson coefficient &   Eigenvalue & Eigenvector \\
\midrule
$c_{E0}$ &    2000  & $0.98 \cdot c_{Hq}^{(3)}$\\
 
$c_{E1}$ &     38       & $0.85 \cdot c_{Hu} - 0.39 \cdot c_{Hq}^{(1)} - 0.27 \cdot c_{Hd}$\\
 
$c_{E2}$ &     8.3        & $0.70 \cdot \Delta\mathrm{BR}/\mathrm{BR}_{\mathrm{SM}} + 0.62 \cdot c_{HW}$\\
 
$c_{E3}$ &    0.2        & $0.74 \cdot c_{HWB} + 0.53 \cdot c_{Hq}^{(1)} - 0.32 \cdot c_{HW}$\\
 
$c_{E4}$ &    $6.4\cdot 10^{-3}$ & $0.65 \cdot c_{HW} - 0.60 \cdot \Delta\mathrm{BR}/\mathrm{BR}_{\mathrm{SM}} + 0.35 \cdot c_{Hq}^{(1)}$ \\
\bottomrule
\end{tabular}
\end{center}
\end{table}
 
\begin{figure}[htbp]
\begin{center}
\includegraphics[width=0.7\linewidth]{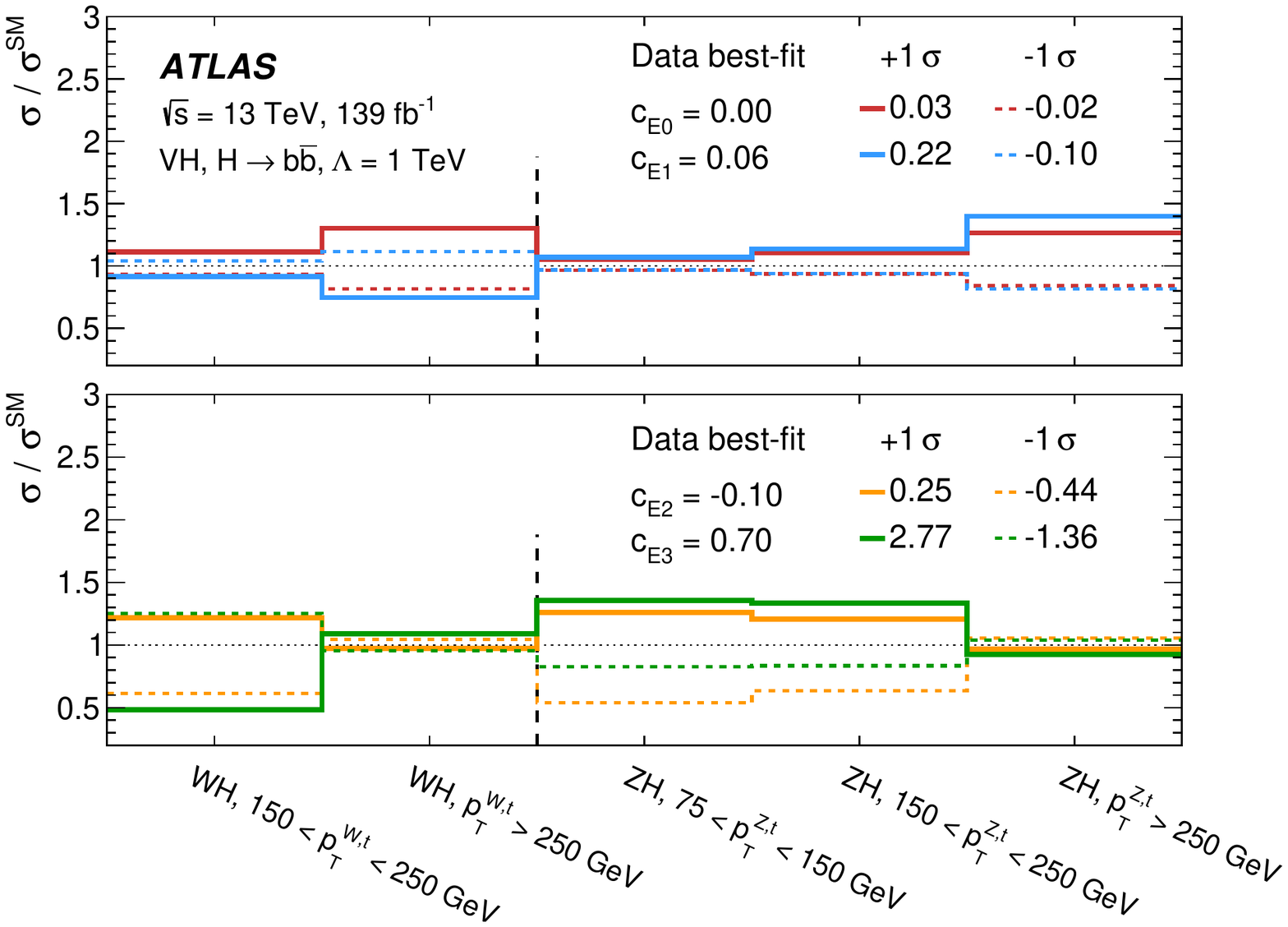}
\end{center}
\caption{The impact of the leading four eigenvectors on the STXS cross-section measurements. The change to the cross-section is indicated at the $+1\sigma$ (solid) and $-1\sigma$ (dashed) limits of the corresponding Wilson coefficients, extracted from a simultaneous fit to data of all five eigenvectors.}
\label{fig:eigenoperator_fit}
\end{figure}
 
\FloatBarrier

\section{Conclusion}
Measurements are presented of the Standard Model Higgs boson decaying into a $b\bar{b}$ pair and produced in
association with a $W$ or $Z$ boson, using data collected by the ATLAS
experiment in proton--proton collisions from Run~2 of the LHC\@. The data
correspond to an integrated luminosity of \lumiFull\ collected at a centre-of-mass energy of $\sqrt{s} = $13~\TeV.
 
For a Higgs boson with  $m_H = 125$~\GeV\ produced in association with either a $Z$ or $W$ boson, an observed (expected) significance of 6.7 (6.7) standard deviations is found and a signal strength relative to the SM prediction of $\mu_{VH}^{bb}~=~1.02^{+0.12}_{-0.11}\mathrm{(stat.)} ^{+0.14}_{-0.13} \mathrm{(syst.)}$ is measured.
For a Higgs boson produced in association with a $W$ boson, an observed (expected) significance of 4.0 (4.1) standard deviations is found and a signal strength relative to the SM prediction for $m_H = 125$~\GeV\ of $\mu_{WH}^{bb}~=~0.95 \pm 0.18 \mathrm{(stat.)} ^{+0.19}_{-0.18} \mathrm{(syst.)}$ is measured.
For a Higgs boson produced in association with a $Z$ boson an observed (expected) significance of rejecting the background-only hypothesis of 5.3 (5.1) standard deviations is found and a signal strength of
$\mu_{ZH}^{bb}~=~1.08 \pm 0.17 \mathrm{(stat.)} ^{+0.18}_{-0.15} \mathrm{(syst.)}$ is measured.
 
Cross-sections of associated production of a Higgs boson decaying into bottom quark pairs and an electroweak gauge boson, $W$ or $Z$, decaying into leptons are measured
as a function of the gauge boson transverse momentum in kinematic fiducial volumes in the simplified template cross-section framework. The uncertainties in the measurements vary from 30\% in the highest \ptv\ regions to 85\% in the lowest, and are in agreement with the Standard Model predictions.
 
Limits are also set on the coefficients of effective Lagrangian operators which affect the $VH$ production and $H \to b\bar{b}$ decay. Limits are studied for both the variation of a single coefficient and also the simultaneous variation of a set of linear combinations of coefficients. The allowed range of the individual or linear combinations of the coefficients, to which the analysis has the greatest sensitivity, is limited to a few percent.
 

\section*{Acknowledgements}

We thank CERN for the very successful operation of the LHC, as well as the
support staff from our institutions without whom ATLAS could not be
operated efficiently.
 
We acknowledge the support of ANPCyT, Argentina; YerPhI, Armenia; ARC, Australia; BMWFW and FWF, Austria; ANAS, Azerbaijan; SSTC, Belarus; CNPq and FAPESP, Brazil; NSERC, NRC and CFI, Canada; CERN; CONICYT, Chile; CAS, MOST and NSFC, China; COLCIENCIAS, Colombia; MSMT CR, MPO CR and VSC CR, Czech Republic; DNRF and DNSRC, Denmark; IN2P3-CNRS and CEA-DRF/IRFU, France; SRNSFG, Georgia; BMBF, HGF and MPG, Germany; GSRT, Greece; RGC and Hong Kong SAR, China; ISF and Benoziyo Center, Israel; INFN, Italy; MEXT and JSPS, Japan; CNRST, Morocco; NWO, Netherlands; RCN, Norway; MNiSW and NCN, Poland; FCT, Portugal; MNE/IFA, Romania; MES of Russia and NRC KI, Russia Federation; JINR; MESTD, Serbia; MSSR, Slovakia; ARRS and MIZ\v{S}, Slovenia; DST/NRF, South Africa; MINECO, Spain; SRC and Wallenberg Foundation, Sweden; SERI, SNSF and Cantons of Bern and Geneva, Switzerland; MOST, Taiwan; TAEK, Turkey; STFC, United Kingdom; DOE and NSF, United States of America. In addition, individual groups and members have received support from BCKDF, CANARIE, Compute Canada and CRC, Canada; ERC, ERDF, Horizon 2020, Marie Sk{\l}odowska-Curie Actions and COST, European Union; Investissements d'Avenir Labex, Investissements d'Avenir Idex and ANR, France; DFG and AvH Foundation, Germany; Herakleitos, Thales and Aristeia programmes co-financed by EU-ESF and the Greek NSRF, Greece; BSF-NSF and GIF, Israel; CERCA Programme Generalitat de Catalunya and PROMETEO Programme Generalitat Valenciana, Spain; G\"{o}ran Gustafssons Stiftelse, Sweden; The Royal Society and Leverhulme Trust, United Kingdom.
 
The crucial computing support from all WLCG partners is acknowledged gratefully, in particular from CERN, the ATLAS Tier-1 facilities at TRIUMF (Canada), NDGF (Denmark, Norway, Sweden), CC-IN2P3 (France), KIT/GridKA (Germany), INFN-CNAF (Italy), NL-T1 (Netherlands), PIC (Spain), ASGC (Taiwan), RAL (UK) and BNL (USA), the Tier-2 facilities worldwide and large non-WLCG resource providers. Major contributors of computing resources are listed in Ref.~\cite{ATL-GEN-PUB-2016-002}.

\clearpage
 
\printbibliography
 
\clearpage
 
 
\begin{flushleft}
\hypersetup{urlcolor=black}
{\Large The ATLAS Collaboration}

\bigskip

\AtlasOrcid[0000-0002-6665-4934]{G.~Aad}$^\textrm{\scriptsize 102}$,    
\AtlasOrcid[0000-0002-5888-2734]{B.~Abbott}$^\textrm{\scriptsize 128}$,    
\AtlasOrcid[0000-0002-7248-3203]{D.C.~Abbott}$^\textrm{\scriptsize 103}$,    
\AtlasOrcid[0000-0002-2788-3822]{A.~Abed~Abud}$^\textrm{\scriptsize 36}$,    
\AtlasOrcid[0000-0002-1002-1652]{K.~Abeling}$^\textrm{\scriptsize 53}$,    
\AtlasOrcid[0000-0002-2987-4006]{D.K.~Abhayasinghe}$^\textrm{\scriptsize 94}$,    
\AtlasOrcid[0000-0002-8496-9294]{S.H.~Abidi}$^\textrm{\scriptsize 167}$,    
\AtlasOrcid[0000-0002-8279-9324]{O.S.~AbouZeid}$^\textrm{\scriptsize 40}$,    
\AtlasOrcid{N.L.~Abraham}$^\textrm{\scriptsize 156}$,    
\AtlasOrcid[0000-0001-5329-6640]{H.~Abramowicz}$^\textrm{\scriptsize 161}$
% ...
\bigskip
\\

% ...
$^{36}$CERN, Geneva; Switzerland.\\
% ...
$^{40}$Niels Bohr Institute, University of Copenhagen, Copenhagen; Denmark.\\
% ...
$^{53}$II. Physikalisches Institut, Georg-August-Universit\"{a}t G\"ottingen, G\"ottingen; Germany.\\
% ...
$^{94}$Department of Physics, Royal Holloway University of London, Egham; United Kingdom.\\
% ...
$^{102}$CPPM, Aix-Marseille Universit\'e, CNRS/IN2P3, Marseille; France.\\
$^{103}$Department of Physics, University of Massachusetts, Amherst MA; United States of America.\\
% ...
$^{128}$Homer L. Dodge Department of Physics and Astronomy, University of Oklahoma, Norman OK; United States of America.\\
% ...
$^{156}$Department of Physics and Astronomy, University of Sussex, Brighton; United Kingdom.\\
% ...
$^{161}$Raymond and Beverly Sackler School of Physics and Astronomy, Tel Aviv University, Tel Aviv; Israel.\\
% ...
$^{167}$Department of Physics, University of Toronto, Toronto ON; Canada.\\
% ...

\end{flushleft}


\clearpage
 
\end{document}